\documentclass[format=acmsmall,screen,anonymous=false, review=false,nonacm]{acmart}

\AtBeginDocument{}

\usepackage{xspace}
\usepackage{listings}
\usepackage{xcolor}
\usepackage{graphicx}
\usepackage{amsmath}

\usepackage{amssymb}

\usepackage{subcaption}
\usepackage{algorithmic}
\usepackage{algorithm}
\usepackage{pifont} 

 \usepackage{enumitem}

\definecolor{codegreen}{rgb}{0,0.6,0}
\definecolor{codegray}{rgb}{0.5,0.5,0.5}
\definecolor{codepurple}{rgb}{0.58,0,0.82}
\definecolor{backcolour}{rgb}{0.95,0.95,0.92}
\lstdefinestyle{TAPstyle}{
	backgroundcolor=\color{backcolour},   
    commentstyle=\color{codegreen},
    keywordstyle=\color{magenta},
    numberstyle=\tiny\color{codegray},
    stringstyle=\color{codepurple},
    basicstyle=\ttfamily\footnotesize,
    breakatwhitespace=false,         
    breaklines=true,                 
    captionpos=b,                    
    keepspaces=false,                 
numbersep=0pt,                  
    showspaces=false,                
    showstringspaces=false,
    showtabs=false,                  
    tabsize=2,
    xleftmargin=5pt,
    xrightmargin=5pt,
morekeywords={rule,when,Item,received,then,end}
}
\usepackage[T1,OT1]{fontenc}

\newcommand{\ie}{\textit{i.e.}\xspace}
\newcommand{\eg}{\textit{e.g.}\xspace}

\newcommand{\mypara}[1]{\noindent\textbf{#1}\xspace}

\newcommand{\code}[1]{{\texttt{#1}}\xspace}
\newcommand{\msg}[1]{{\texttt{#1}}\xspace}

\newcommand{\mycomment}[1]{}
\newcommand{\FIGREF}{Fig.\xspace}
\newcommand{\FIGLOC}{.}

\renewcommand{\algorithmiccomment}[1]{/*~\code{#1}~*/ }

\newcommand{\oh}{OpenHAB\xspace}
\newcommand{\smt}{SmartThings\xspace}
\newcommand{\ifttt}{IFTTT\xspace}

\newcommand{\expat}{{\texttt{{ExPAT}}}\xspace}
\newcommand{\iotguard}{{\texttt{{IoTGuard}}}\xspace}
\newcommand{\maverick}{{\texttt{{Maverick}}}\xspace}
\newcommand{\patriot}{{\texttt{{PatrIoT}}}\xspace}
\newcommand{\helion}{{H$\epsilon$lion}\xspace}
\newcommand{\sysname}{{{\textsc{VetIoT}}}\xspace}

\newcommand*\circledOne{\ding{202}}
\newcommand*\circledTwo{\ding{203}}
\newcommand*\circledThree{\ding{204}}
\newcommand*\circledFour{\ding{205}}
\newcommand*\circledFive{\ding{206}}
\newcommand*\circledSix{\ding{207}}

\newcommand{\iotsystem}{\ensuremath{\mathbb{I}}\xspace}
\newcommand{\States}{\ensuremath{\mathcal{S}}\xspace}
\newcommand{\Actions}{\ensuremath{\mathcal{A}}\xspace}
\newcommand{\Vars}{\ensuremath{\mathcal{V}}\xspace}
\newcommand{\Updates}{\ensuremath{\Lambda}\xspace}
\newcommand{\update}{\ensuremath{\lambda}\xspace}
\newcommand{\Rules}{\ensuremath{\mathcal{R}}\xspace}
\newcommand{\Transitions}{\ensuremath{\mathcal{T}}\xspace}
\newcommand{\Events}{\ensuremath{\mathcal{E}}\xspace}
\newcommand{\Conds}{\ensuremath{\mathcal{C}}\xspace}
\newcommand{\domain}{\ensuremath{\mathcal{D}}\xspace}

\newcommand{\Policies}{\ensuremath{{\Psi}}\xspace}
\newcommand{\Policy}{\ensuremath{{\psi}}\xspace}
\newcommand{\Mapper}{\ensuremath{{\aleph}}\xspace}

\newcommand{\decisionFunction}{\ensuremath{{\Delta}}\xspace}

\newcommand{\TDS}{\ensuremath{\mathbb{T}}\xspace}
\newcommand{\TB}{\ensuremath{\Omega}\xspace}
\newcommand{\TSc}{\ensuremath{\Gamma}\xspace}
\newcommand{\EventSeq}{\ensuremath{\zeta}\xspace}

\newcommand{\Oracle}{\ensuremath{\upsilon}\xspace}

\newcommand{\success}{\texttt{success}\xspace}
\newcommand{\failure}{\texttt{failure}\xspace}

\newcommand{\degree}{\ensuremath{^\circ}\xspace}

\renewcommand\appendix{\par
	\setcounter{section}{0}
	\setcounter{subsection}{0}
	\setcounter{figure}{0}
\renewcommand\thesection{\Alph{section}}
	\renewcommand\thefigure{\Alph{section}\arabic{figure}}
\renewcommand\theHsection{\thesection}
	\renewcommand\theHsubsection{\thesubsection}
}


\begin{document} 

\title[\sysname]{\sysname: On Vetting IoT Defenses Enforcing Policies at Runtime}
\titlenote{A preliminary version of this paper was presented at the conference IEEE CNS 2023 \cite{vetiotCNS2023}.
	This version has been extended with significant additions, such as new features and expanded evaluation results. 
}

\author[AJ Nafis]{Akib Jawad Nafis}
\affiliation{
	\department{EECS}
	\institution{Syracuse University}
	\city{Syracuse}
	\state{NY}
	\country{USA}
}
\email{anafis@syr.edu}
\author[S Hasan]{S Mahmudul Hasan}
\affiliation{
	\department{EECS}
	\institution{Syracuse University}
	\city{Syracuse}
	\state{NY}
	\country{USA}
}
\email{shasan@syr.edu}
\author[O Chowdhury]{Omar Chowdhury}
\affiliation{
	\department{EECS}
	\institution{Stony Brook University}
	\city{Long Island}
	\state{NY}
	\country{USA}
}
\email{omar@cs.stonybrook.edu}
\author[E Hoque]{Endadul Hoque}
\affiliation{
	\department{EECS}
	\institution{Syracuse University}
	\city{Syracuse}
	\state{NY}
	\country{USA}
}
\email{enhoque@syr.edu}

\begin{abstract}
	Smart homes, powered by programmable IoT platforms, often face safety and security issues. A class of defense solutions dynamically enforces policies that capture the expected behavior of the IoT system. Despite numerous innovations, these solutions are under-vetted. The primary reason lies in their evaluation approach---they are self-assessed in isolated virtual testbeds with hand-crafted orchestrated scenarios that require manual interactions using the platform's user-interface (UI). Such non-uniform evaluation setups limit reproducibility and comparative analysis. Closing this gap in the traditional way requires a significant upfront manual effort, causing researchers to turn away from  large-scale comparative empirical evaluation. To address this, we propose VetIoT---a highly automated, uniform evaluation platform---to vet the defense solutions that hinge on runtime policy enforcement. Given a defense solution, VetIoT readily instantiates a virtual testbed to deploy and evaluate the solution. VetIoT replaces manual UI-based interactions with an automated event simulator and manual inspection of test outcomes with an automated comparator. VetIoT incorporates automated event generators to feed events to the event simulator. We developed a prototype of VetIoT, which successfully reproduced and comparatively assessed four runtime policy enforcement solutions. VetIoT's stress testing and differential testing capabilities make it a promising tool for future research and evaluation.
\end{abstract}


\settopmatter{printacmref=false}
\setcopyright{none}
\renewcommand\footnotetextcopyrightpermission[1]{}

\maketitle

\section{Introduction}
\label{sec:intro}

Smart 
homes are powered by numerous programmable IoT platforms (\eg, \smt, \oh, \ifttt)
to facilitate automation and over-the-air control and monitoring. Despite tremendous innovations,
smart home systems suffer from safety and security issues
\cite{fernandes2016smart, wang2019charting}.
As a result, this problem has garnered much attention from the security
research community, which has led to several proposals for defense solutions \cite{ contexiot17, wang2018fear,  iotsafety2019,  iotguard2019ndss, yahyazadeh2019expat, yahyazadeh2020patriot,  chi2020crossapp, ding2021iotsafe,mazhar2023maverick} focused on 
curbing unexpected (\ie, unsafe/insecure) behavior on these platforms.
One category of these defenses
\cite{iotguard2019ndss, yahyazadeh2019expat, yahyazadeh2020patriot,ding2021iotsafe,mazhar2023maverick}
hinges on the enforcement of safety and security policies at runtime,
where each policy essentially captures the expected behavior of 
the IoT system regarding one or more installed IoT devices.
Unfortunately, they are all under-vetted because they were evaluated against a 
small number of testcases. 
In addition, identifying a better defense solution has been difficult as they all lack comparative empirical evaluation.

The reason behind using a small number of testcases lies in their
current evaluation methodology. First, a testbed (predominantly, virtual)
along with the desired automation apps and policies 
is manually set up to evaluate the defense solution in isolation. 
Secondly, an evaluator comes up with testcases, which have to be
manually orchestrated. A testcase defines a sequence of events that can occur 
naturally (\eg, a certain time of the day) or due to an interaction with a IoT device 
(\eg, opening the front door). The evaluator feeds a testcase into the testbed
by manually interacting with the platform's UI (\eg, a web-based UI or a phone app). 
The testcases that can demonstrate the defense's efficacy are primarily selected for evaluation. 
Thirdly, after feeding a testcase, the log information from the platform is collected and manually inspected
for reporting the achieved efficacy.
Finally, for the next testcase, a manual cleanup of the testbed is required to bring it back 
to the initial (clean) state. 
This huge upfront manual effort causes the researchers to resort to only hand-crafted testcases
for evaluation, leaving the defense solutions mostly under-vetted.

Such non-uniform evaluation setups explain 
the absence of comparative empirical assessments of the effectiveness of these defense solutions.
For comparing a new defense (say, $\mathcal{A}$) with an existing one ($\mathcal{B}$), 
an evaluator must be able to test both $\mathcal{A}$ and $\mathcal{B}$ 
on the $\mathcal{B}$'s original test-suite and a new test-suite. 
The former calls for reproducing $\mathcal{B}$'s empirical results, 
and the latter calls for a uniform evaluation mechanism irrespective 
of the differences between $\mathcal{A}$ and $\mathcal{B}$. 
Unfortunately, the traditional way of evaluating these defense solutions
limit both reproducibility and comparative analysis.

As an example, consider a smart home testbed where the initial states of the devices
are: $\lbrace$\code{IndoorMotionSensor = OFF}, \code{FrontDoor = CLOSED}, \code{HomeMode = ON}, \code{SleepMode = OFF}
$\rbrace$.
In addition, an automation app ``\code{R1: When IndoorMotionSensor senses motion,} \code{execute FrontDoor.Open()}''
and a policy ``\code{P1: Deny opening FrontDoor,} \code{if user is not Home}'' are installed in the testbed.
Suppose we have a testcase (aka, a sequence of events): 
\code{$\langle$HomeMode = OFF, IndoorMotionSensor = ON$\rangle$}. 
The first event simply changes the state of \code{HomeMode} to \code{OFF}, and
the second event triggers \code{R1}, which attempts to open \code{FrontDoor}.
While defenses like \expat \cite{yahyazadeh2019expat} and \patriot \cite{yahyazadeh2020patriot} block 
\code{FrontDoor.Open()} as opening \code{FrontDoor} when the user is away will violate \code{P1}, 
the other solutions such as \iotguard \cite{iotguard2019ndss} and IoTSafe \cite{ding2021iotsafe}
fail to block \code{FrontDoor.Open()} even though the user is away. Without testing against a large number of testcases 
and evaluating for a comparative analysis, 
the shortcomings of a defense solution will remain undetected.

Prior efforts on some standardization in IoT evaluations
proposed testbeds with physical devices \cite{munoz:hal-02266558,285062}
or emulated devices \cite{10049670}, and benchmarks consisting of IoT applications \cite{almakhdhub2019benchiot,iotbench}. 
However, none of these efforts automate the evaluation of 
policy enforcing IoT defense solutions.

In this paper, we present \sysname, a highly-automated uniform evaluation
platform for vetting smart home defenses that hinge on runtime policy
enforcement. \sysname is designed to address the aforementioned limitations
by automating much of the experimenting process.
First, \sysname incorporates a testbed generator to quickly and easily create an identical testbed to enable a controlled vetting environment 
required for reproducibility and comparative testing.
Secondly, 
\sysname employs an event sequence generator to automate the generation of testcases. 
Instead of manually feeding each event using the platform's UI, \sysname utilizes
the external programming interface available on the platform 
(\eg, REST API) to push each event of the testcase. Finally, \sysname replaces manual inspection of the test outcomes with an automated comparator
to report on the defense solution's efficacy.
The comparator makes decisions based on a heuristic algorithm that checks for discrepancy between the initial and final states of the testbed.

We developed a fully functional prototype of \sysname using Python 3.9. 
\sysname leverages virtual IoT devices
as they cost nothing and require less time to reset and boot up 
as opposed to physical devices.
To deploy the testbed, \sysname utilizes the \oh IoT platform \cite{OpenHab},
because it is not only open source but also entirely deployable on a local machine, as opposed to 
\smt \cite{smartThings} which is hosted on a proprietary cloud server.
When \sysname interacts with the testbed (\eg, to push a testcase), 
it uses \oh's REST API interface. 

We vetted four smart home defense solutions that
enforce safety and security policies at runtime: 
\patriot \cite{yahyazadeh2020patriot}, \expat \cite{yahyazadeh2019expat}, \iotguard \cite{iotguard2019ndss}, and \maverick \cite{mazhar2023maverick}.
We selected them for vetting because of their
popularity in the research community.
\expat,\footnote{\url{https://github.com/expat-paper/expat}}
\patriot,\footnote{\url{https://github.com/yahyazadeh/patriot}}
and \maverick \footnote{\url{https://github.com/hammadmazhar1/MAVERICK}}
have open-source implementations for \oh available on GitHub.com.
\iotguard,\footnote{\url{https://github.com/Beerkay/SmartAppAnalysis}}
on the other hand, is designed for \smt, but
its public repository on GitHub.com does not contain the full implementation,
which prompted us to develop an in-house implementation of \iotguard for \oh.
Our implementation of \iotguard
faithfully followed its description \cite{iotguard2019ndss} as closely 
as possible.

Using \sysname, we evaluated each test-subject (\expat, \patriot, \iotguard, \maverick) 
by following three testing approaches: \textit{fidelity}, \textit{stress}, and \textit{differential}.
Technically, fidelity testings are used to assess \sysname's fidelity in reproducing 
the efficacy results of each subject by replicating the same testbed and testcases as mentioned in 
\cite{yahyazadeh2019expat,yahyazadeh2020patriot, iotguard2019ndss, mazhar2023maverick}. 
Next, we used \sysname to stress test each subject separately to measure their own efficacy with respect to multiple test-suites comprising a total of 280 testcases.
\sysname incorporates a random and a language model-based event generators to produce these testcases.
The random generator simply produces testcases by randomly generating events plausible for the devices
of the testbed.
The language model-based event generator, proposed by \helion \cite{manandhar2020Helion},
produces testcases with realistic smart home scenarios. 
Finally, \sysname was used to perform comparative analysis 
of \expat, \patriot, \iotguard, and \maverick via a differential testing,
where we used an identical testbed along with the same automatically generated testcases.
We performed differential testing on 6 different testbeds.
Our vetting revealed many nuanced corner cases where the efficacy of a test subject differed
for unique reasons specific to their design choices (\eg, which policies are selected to enforce at runtime). 
\sysname is available as open-source at \code{https://github.com/syne-lab/vetiot}.
{A preliminary version of this paper was presented at the
	conference IEEE CNS 2023 \cite{vetiotCNS2023}.}

\mypara{Contributions.} This paper makes the following contributions: 
\begin{itemize} \item We proposed a highly-automated uniform evaluation platform, dubbed \sysname, for vetting IoT defenses
	that enforce policies at runtime. It automates much of the traditional evaluation process, which 
	requires huge manual efforts from the researchers.

	\item We developed a fully functional prototype of \sysname for the \oh platform. 
	To demonstrate \sysname's efficacy, we evaluated \sysname by applying it to four IoT defenses 
	(\expat, \patriot, \iotguard, and \maverick) for assessing
	their individual and comparative efficacy.
	We enabled evaluation of IoT defenses with natural smart home scenarios
	by incorporating \helion into \sysname.
	
	\item To the best of our knowledge, \sysname is the first automated platform that empirically evaluates 
	IoT security defenses enforcing policies at runtime. 
\end{itemize}

The rest of the paper is organized as follows. Section~\ref{sec:prelim} provides the necessary 
backgrounds. Section~\ref{sec:overview} and \ref{sec:design} describe an overview and 
the design of \sysname, respectively. \sysname's implementation and evaluation are presented in 
Section~\ref{sec:impl} and \ref{sec:eval} followed by a discussion in Section~\ref{sec:discuss} 
and the related work in Section~\ref{sec:related}. Section~\ref{sec:conclusion} summarizes the paper.

\section{Preliminaries}\label{sec:prelim}

\mypara{Platforms.}
Smart home platforms (\eg, \oh, \smt) provide users with a common interface to control and manage IoT devices
and enable users to automate numerous manual tasks with customized applications (aka, \textit{apps}) 
that may interact with the physical world by operating the IoT devices. 
A platform is considered as the \textit{brain} of a smart home, because 
the platform not only manages the devices connected to it and ensures communication between them
but also executes the core automation logic. Depending on the platform's architecture, 
a local or remote server (aka, \textit{backend}) is employed to host the brain. For instance, \smt utilizes a proprietary cloud-based backend whereas
\oh allows users to spawn their own local backend.

\mypara{Apps.}
Platforms typically allow numerous customization in automation apps, which can vary in complexity. 
For example, ``\code{When the sun sets, turn on the porch lights}'' is a simpler app and 
``\code{When Smoke\_Detector detects smoke 
	and} \code{if Living\_room\_temperature > 120\degree{},} \code{turn on Fire\_Sprinkler}'' is a 
complex one. In general, these apps
follow a \textit{trigger-condition-action} paradigm.
A trigger, which is usually a logical event occurred in the smart home (\eg, ``\code{Smoke\_Detector detects smoke}''), 
initiates the execution of an app. 
The action block includes a list of commands to be operated on the respective IoT device (\eg, ``\code{turn on Fire\_Sprinkler}''). 
Note that a command can change not only the state of the physical world (\eg, turning on the fire sprinkler will spray water) 
but also the internal state of the device (\eg, \code{Fire\_Sprinkler.state = ON}).
A condition is an optional block of predicates (\eg, \code{Living\_room\_temperature > 120\degree{}}), 
often expressing a situation that
must be satisfied before the app can continue executing its action block.

\mypara{Defenses using Policy Enforcement.}
Many safety and security concerns of smart home (or, IoT systems in general) often stem from faulty apps or 
unintended interaction and interference between apparently correctly functioning apps.
It is a common threat model for smart homes.
To curb the undesired behavior of apps, many solutions have been proposed, including 
a class of defenses that rely on enforcing policies at runtime to ensure safety 
and security of IoT systems \cite{iotguard2019ndss, yahyazadeh2019expat, yahyazadeh2020patriot,ding2021iotsafe,mazhar2023maverick}.  
A policy is basically a user's expectation about the behavior of an IoT system. 
An example policy can be ``\code{The surveillance camera can never be turned off}.''
For example, if an app issues an action to turn off the surveillance camera (\ie, \code{camera.off()} command), 
a defense solution should block it as the action violates the camera policy.
Thus, the defense attempts to keep the system aligned with the user's expectations at runtime.

Furthermore, for policy enforcement, a defense solution needs to deploy hooks (called policy information point (\textbf{PIP})) 
at several locations in each app
to collect additional information at runtime, necessary to decide about policies using 
a policy decision function (\textbf{PDF}). 
The defense also uses 
a  policy enforcement point (\textbf{PEP}) where 
the defense checks if the contemplated
actions will satisfy or violate the user-provided policies.

Where to deploy their PEP, PIP and PDF varies among the defense mechanisms.
While some solutions \cite{yahyazadeh2019expat,yahyazadeh2020patriot} deploy all its PEP, 
PIP and PDF in the app's source code, solutions like \cite{iotguard2019ndss,ding2021iotsafe}
keep PIP and PEP in the app but 
leverage a remote offshore server to deploy their PDF. 
Therefore, they \cite{iotguard2019ndss,ding2021iotsafe} also need an access to
the apps' source code for instrumentation.

Irrespective of how PEP, PIP, and PDF are deployed, 
whenever the execution of an app reaches its PEP (\eg, its action block),
the instrumented code in PEP invokes its PDF by supplying all the information collected from its PIP.
Once the PDF returns with a response, the PEP simply allows the actions if the response is positive 
or blocks the actions if the response is negative.

A recent category of defense like \maverick \cite{mazhar2023maverick} 
deploys all its components (\ie, PIP, PDF, PEP) in a trusted intermediary node (aka, a proxy) that
is on-path between IoT devices and the platform's backend. The proxy intercepts and monitors all smart home communications, 
such as events and actions, exchanged between IoT devices and the platform.
Therefore, defenses like \maverick do not require any instrumentation of IoT apps.
\maverick defends the system by enforcing policies at the proxy as follows.
\maverick blocks an intercepted action---issued either by an IoT app running on the platform or by a third-party 
service---only if the action would violate a policy; otherwise, \maverick allows the action. Similarly, \maverick 
takes a predefined corrective action whenever it finds that an intercepted event from an IoT device has violated
a policy.

\section{Overview of \sysname}
\label{sec:overview}

In this section, we present an abstract model of our testing platform and our problem definition.

\subsection{Abstract Models}
\label{sec:model}

\noindent
\textbf{Programmable IoT Systems.}
At a high-level, a programmable IoT system \iotsystem can be viewed a labeled transition system (LTS) defined
as $\iotsystem = \langle \Vars, \States, \Actions, \Updates, \Rules, \Transitions \rangle$. 
Here, \Vars represents a finite set of \textit{typed} variables, which can further be
decomposed into two mutually exclusive sets $\Vars_{env}$ and $\Vars_{dev}$.  
$\Vars_{env}$ refers to a set of environment variables (\eg, temperatures) and $\Vars_{dev}$
denotes a set consisting of the internal state variables of each device deployed in this system 
(\eg, the state of a bedroom light).
\States is a non-empty finite set of system-states such that each $s \in \States$ is 
a tuple $\langle d_0, d_1, \ldots, d_{|\Vars|} \rangle$ where $d_i$ is an assigned value to $v_i \in \Vars$ taken from $v_i$'s finite domain $\domain_i$.
For example, a state $s$ of an IoT system composed of two devices -- a front-porch light and a smart lock --
can be $\langle \code{On, Locked} \rangle$ at a given instant of time. 
\Actions in \iotsystem refers to a finite set of all possible activities supported by the underlying IoT
platform. Technically, an IoT platform provides a list of \code{action\_command}{}s to change 
each $v \in \Vars$. For example, 
$\{\code{Lock, Unlock}\}$ are two possible actions allowed to change the status of a smart lock. 

The rest of the components of \iotsystem is relatively complex.
Whenever there is a change in the value of $v_i$ (\eg, from $d_{i_j}$ to ${d_{i_k}}$, 
where $d_{i_j}, {d_{i_k}} \in \domain_i$), the device/sensor 
sends $\update_{i}$ (aka, a \msg{status\_update} message) to the backend. In fact, there is a one-to-one mapping
between $v_{i}$ and $\update_{i}$, and hence $\Updates = \bigcup_{i=0}^{n} \update_i$. 
The change in $v_i$ can happen \textit{after} (i) the device executes an appropriate command 
$a_x \in \Actions$
received from the backend (\eg, the \msg{Unlock} command), 
or (ii) a user physically interacts with the device (\eg, unlocking the door with a key).

\Rules defines a set of all possible automation apps allowed by the underlying platform and can be viewed as 
$\Rules \subseteq \Events \times \Conds \times 2^{\Actions}$. 
Here, \Events is a finite set of all possible methods to interact with \iotsystem. These
interactions can be either through physical interactions with the actual device(s) 
or using the platform's UI. The former method causes the device to send $\update_i$ to the backend
for the change occurred in $v_i$, 
while the latter method asks the backend to send $a_{x}$ to the device which may change $v_i$. 
Hence, \Events is essentially the possible set of \textbf{triggering events}, defined as $\Events = \Actions \cup \Updates$
(Recall that \Actions and \Updates are mutually exclusive. Some platforms, \eg, \oh, allow
apps that can be triggered right before executing an action from \Actions).
\Conds is a finite set of conditions/predicates such
that each element $c \in \Conds$ is 
a Boolean expression with logical and relational operators over \Vars. 
In other words, a condition dictates a specific situation under which the app in question can 
be executed. 
For any $e \in \Events$, $c \in \Conds$ and $\alpha \in 2^{\Actions}$, 
if an app $\langle e, c, \alpha \rangle \in \Rules$, then it signifies 
that after observing a triggering event $e$ under a situation where $c$ evaluates to true, 
the backend will execute the actions listed in $\alpha$.

\Transitions is another relation and defined as 
$\Transitions \subseteq \States \times \Events \times \States$. 
This signifies how the system \iotsystem transitions from one state to another upon 
receiving a triggering event from a device. We consider \Transitions to be left-total
and \iotsystem to be a deterministic LTS.

\noindent
\textbf{Target Defense Solutions (TDS).} In this paper, our target defense solutions
are those that aspire to ensure the safety and security of an IoT system
by enforcing policies at runtime \cite{yahyazadeh2019expat,yahyazadeh2020patriot,iotguard2019ndss, mazhar2023maverick}.
A target defense system (TDS) \TDS is defined as
$\TDS = \langle \Policies, \decisionFunction, \Mapper \rangle$.  
Here, \Policies is a finite set of policy statements (say, $\Policy_1, \ldots, \Policy_n$), 
\decisionFunction dictates the logic behind its policy enforcement (\ie, PIP, PEP, and PDF), and \Mapper is a special function to
embed PEP, PIP and PDF in each app before installing it to \iotsystem. 

Assume for each app $r \in \Rules$, $\xi(r)$ denotes the syntactical representation of $r$ 
written in the language supported by the platform
and $\mathcal{L}(r)$ denotes the semantics  of $r$ defined by the language's type system and the supported logical formula. 
Now we can defined \Mapper, which is
in theory, $\Mapper: \Rules \mapsto \Rules$. In other words, for each $r \in \Rules$,
there exists $r' \in \Rules$ such that both $r$ and $r'$ are semantically the same (\ie, $\mathcal{L}(r) \equiv \mathcal{L}(r')$)
but syntactically different (\ie, $\xi(r) \neq \xi(r')$). The syntactical difference between $r$ and $r'$ 
is merely due to incorporating  PIP, PEP, and PDF inside $r'$.

The target system is often equipped with its own domain specific language 
that the user utilizes to write each policy. Internally, each policy $\Policy_j$
is first converted to a logical formula (preferably, quantifier-free first-order logic (QF-FOL) or similar). 
In most cases, the collection of policies \Policies is treated as a logical 
conjunctive formula over all $\Policy_j${}s (\ie, $\Policies \equiv \bigwedge_{j=0}^{n} \Policy_j$).

The policy decision logic \decisionFunction expresses the underlying 
mechanism to enforce \Policies at runtime. This mechanism varies with each TDS. 
For example, given \Policies, $\decisionFunction_{\expat}$ used by \expat \cite{yahyazadeh2019expat}
influences $\Transitions$ to include 
$\langle s_i, \update, s_j \rangle$
only if
$s_j \models \Policies$. 
Informally, $\decisionFunction_{\expat}$ will allow a contemplated action $a \in \Actions $
in $s_i$ only
if the status update $\update$ that will be 
resulted after executing $a$ will take the system to $s_j$, 
which satisfies \Policies (\ie, $s_2 \models \Policies $, aka, safe). 
$\decisionFunction_{\iotguard}$ used by 
\iotguard \cite{iotguard2019ndss} logically achieves the same outcome 
as $\decisionFunction_{\expat}$.
On the contrary, given \Policies,  $\decisionFunction_{\patriot}$ of \patriot \cite{yahyazadeh2020patriot} 
allows $a \in \Actions$ in $s_i$ only if 
$s_i \models \Policies$ and thus includes 
$\langle s_i, a, s_j \rangle \in \Transitions$.
Informally, $\decisionFunction_{\patriot}$ allows $a$ only if
the current state $s_i$ satisfies all policies relevant to $a$.

\noindent
\textbf{Testbeds:}
A testbed is defined as $\TB = \langle \iotsystem, \TDS \rangle$, where 
\iotsystem and \TDS are defined as above. For example, when \sysname 
instantiates a testbed for \expat, the testbed is notified as 
$\TB_{\expat} = \langle \iotsystem_{\expat}, \TDS_{\expat} \rangle$. Similarly, for 
\patriot, \iotguard, and \maverick , \sysname can instantiate 
$\TB_{\patriot} = \langle \iotsystem_{\patriot}, \TDS_{\patriot}\rangle $,
$\TB_{\iotguard} = \langle \iotsystem_{\iotguard}, \TDS_{\iotguard}\rangle $, 
and $\TB_{\maverick} = \langle \iotsystem_{\maverick}, \TDS_{\maverick}\rangle $ respectively.

Note it is possible that the testbeds can 
have the same \iotsystem (\ie, $\iotsystem_{\expat} = \iotsystem_{\patriot} = \iotsystem_{\iotguard} = \iotsystem_{\maverick}$), 
but their \TDS{s} must be different.
It is also possible that the IoT system (\iotsystem) of the testbed comes from a 
community dataset, such as \helion \cite{manandhar2020Helion}.
For example, \sysname can instantiate a testbed for \expat 
with a smart home (\iotsystem) based on the \helion's description, which will be notified as $\iotsystem_{Helion\_x}$. 
Finally, when no target defense solution is selected, the testbed 
$\TB_{\Oracle} = \langle \iotsystem, \varnothing \rangle$
is considered as a \textbf{vanilla} testbed, 
where neither instrumentation on apps nor policy enforcement at runtime are performed.

\noindent
\textbf{Testcases:}
Given a testbed $\TB$ (an instantiation for a \iotsystem and \TDS), a testcase (aka, test scenario) is defined as 
$ \TSc = \langle \TB, \EventSeq \rangle $, where \EventSeq is a sequence
of triggering events (recall, \Events).

\subsection{Problem Definition}
\label{sec:prob-def}

Consider a testcase $ \TSc = \langle \TB_x, \EventSeq \rangle $, where
$\TB_x$ be a testbed instantiated for a $\iotsystem_x$ and an actual $\TDS_x$ and 
\EventSeq be $\langle e_{1}, e_2, \ldots, e_n \rangle $
with $n > 0$. 
Assume \iotsystem begins at $s_0$. After pushing each $e_j$ (where, $1 \le j \le n$)
to $\TB_x$, $\iotsystem_x$ transitions through states $s_j$ and eventually ends up 
in $s_n$. 
If any $e_j \in \EventSeq$ triggers the app that in turn invokes some unexpected actions, 
\sysname can detect if $\TDS_x$ were successful in preventing the unexpected 
actions and mark \TSc accordingly (\ie, \success or \failure).
\success signifies that $\TDS_x$ prevented the unexpected actions whereas 
\failure denotes $\TDS_x$ failed to do so.

\mypara{Problem.}
\textit{Given a TDS (say, $P$), can \sysname evaluate its efficacy?}
We designed \sysname to evaluate the TDS using three testing approaches:

\begin{itemize}[topsep=0pt,itemsep=2pt] \item \textit{Fidelity testing}: \sysname checks if it can reproduce the evaluation result reported 
	in the TDS's paper/repository. Formally,
	\sysname instantiates $\TB_P$ using the same $\iotsystem_P$ and $\TDS_P$ as described in $P$'s paper
	and tests the TDS against a series of testcases 
	$\TSc_1, \TSc_2, \ldots$, where each $\TSc_i  = \langle \TB_P, \EventSeq_i \rangle$ and 
	$\EventSeq_i$ be the same event sequence used in the paper.
	
	\item \textit{Stress testing}: \sysname measures the TDS's efficacy against 
	new testcases (not tested in the original paper). Formally, \sysname utilizes
	the previously instantiated $\TB_P$ and tests against a series of testcases 
	$\TSc_1, \TSc_2, \ldots$, where each $\TSc_i  = \langle \TB_P, \EventSeq_i \rangle$ and 
	$\EventSeq_i$ be an event sequence generated automatically.
	
	\item \textit{Differential testing}: \sysname assesses how the given TDS ($P$) fares against
	other TDSs (say, $Q$). Formally, \sysname first 
	instantiates two almost identical testbeds $\TB_P$ and $\TB_Q$ where
	$\iotsystem_P = \iotsystem_Q$ and $\Policies_P = \Policies_Q$ and then tests both $P$ and $Q$ in their 
	respective testbed using the same series of automatically generated event sequences. 
	Finally, \sysname reports the efficacy of $P$ compared to $Q$.
	 
\end{itemize}

\section{Design of \sysname}
\label{sec:design}

We will describe a high-level workflow of \sysname followed by the inner workings of its different components.

\subsection{Workflow}
\label{subsec:SystemWorkflow}
\FIGREF~\ref{fig:sys-arch} presents the architecture of \sysname.  
\sysname consists of 4 modules: Testbed Generator, Event Sequence Generator, Event Simulator, and Comparator. 
All four modules of the \sysname work in concert to create a desired testbed and conduct experiments for
evaluating an IoT defense solution.

Given a testbed configuration file provided by the user (aka, the evaluator), 
the testbed generator instantiates a virtual testbed -- consisting of IoT devices -- 
in a programmable IoT platform
(\eg, \oh) and then installs the specified automation apps. For the vanilla testbed $\TB_{\Oracle}$, 
that is all required. However, to prepare a testbed $\TB_{\TDS}$ for a target defense solution (\TDS), the testbed generator deploys the supplied policies in appropriate format and the solution itself.
A testbed is instantiated automatically by invoking (external) APIs of the platform (\circledOne).
Our generator can create not only an arbitrary testbed but also a customized testbed based on the supplied
configuration. While the former mode offers versatility, the latter enables a controlled vetting environment essential for reproducibility and comparative analysis.

After that, the event simulator takes over the control and drives the execution of each testcase. 
Upon the request for the next testcase (\circledTwo), the event sequence generator
supplies a sequence of events ($E_i = \langle e_1, \ldots, e_n \rangle$) 
composing the testcase $\TSc_i$ (\circledThree).\footnote{$E$ and $\EventSeq$ are used interchangeably to denote a sequence of events.}
What an event will be depends on the devices installed in the testbed. For example, 
\code{TV = ON} is a possible event for \code{TV}, indicating \code{TV} has been turned \code{ON};
similarly, \code{MotionSensor = ON} is possible event for the \code{MotionSensor} device, indicating
the motion sensor has detected motion.
Now whether the supplied sequence of events 
is selected from a predefined set or automatically generated 
depends on the configuration provided to \sysname.
The former is used for fidelity testing of each TDS 
and the latter is used for stress and differential testing.

To generate events automatically, \sysname incorporates two generators---either can be selected
through configuration.
The first generator randomly produces events that are plausible based on the devices
of the testbed.
On the contrary,
the second generator, proposed by \helion \cite{manandhar2020Helion},
aims to produce more realistic smart home events. This generator internally relies on
a language model that was trained on IoT applications collected by surveying several smart home users.
(More on this later in $\S$~\ref{sec:event-gens}.)

The event simulator pushes each event $e_j$ sequentially to the platform using the platform's external API
(\circledFour). 
Upon processing each event $e_j$, the testbed logically changes from 
the system-state $s_{j-1}$ to $s_j$. Note that $s_{j-1}$ and $s_j$ are not necessarily unique.
After a short delay, the comparator collects the testbed's initial $s_0$ and final $s_n$ system-states (\circledFive)
and compares them to generate a report (\circledSix). This report indicates
whether the \TDS was successful in preventing unexpected actions in case of an event causing any policy violation.
\sysname additionally provides a debug mode. If it is enabled, a trace of how the system-state is changing
after each event is included in the report.
Before the event simulator repeats the process for the next testcase from (\circledTwo), 
\sysname automatically resets 
the testbed to its initial. The testing cycle will terminate once there is
no more testcases or the time budget expires.

\begin{figure}[!t]
	\centering
	\includegraphics[width=0.7\textwidth]{\FIGLOC/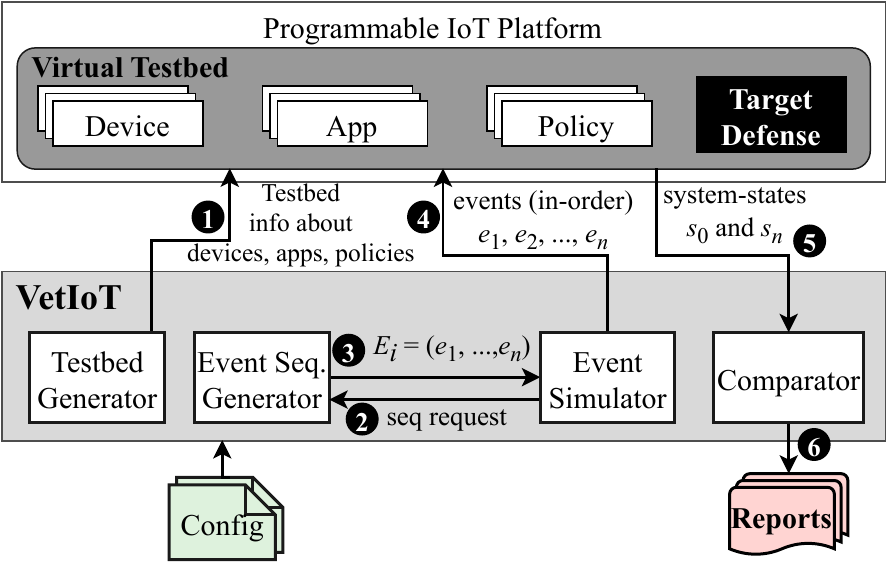}
	\caption{\sysname's architecture and workflow}
	\label{fig:sys-arch}
\end{figure}

\subsection{Automated Comparator}
\label{sec:applyVetIoT}
The goal of designing \sysname is to evaluate dynamic policy enforcing smart home defenses automatically.
To perform automated testing, \sysname easily replaces most of the manual steps of the traditional evaluation: 
setting up testbed, generating testcases, and running testcases. 
But automating the manual inspection of test outcome is challenging, 
because neither the platform nor a TDS itself provides any output signal 
to indicate the success/failure of the TDS in preventing unexpected actions.
One can argue to export the internal result of policy violation from inside the platform or the TDS, 
which we want to avoid as it will require us to modify either the platform or the TDS, limiting \sysname's portability.
With these constraints, there exists no unequivocal way to externally measure the efficacy of the TDS.

\begin{figure}[!t]
	\centering
	\includegraphics[width=0.6\textwidth]{\FIGLOC/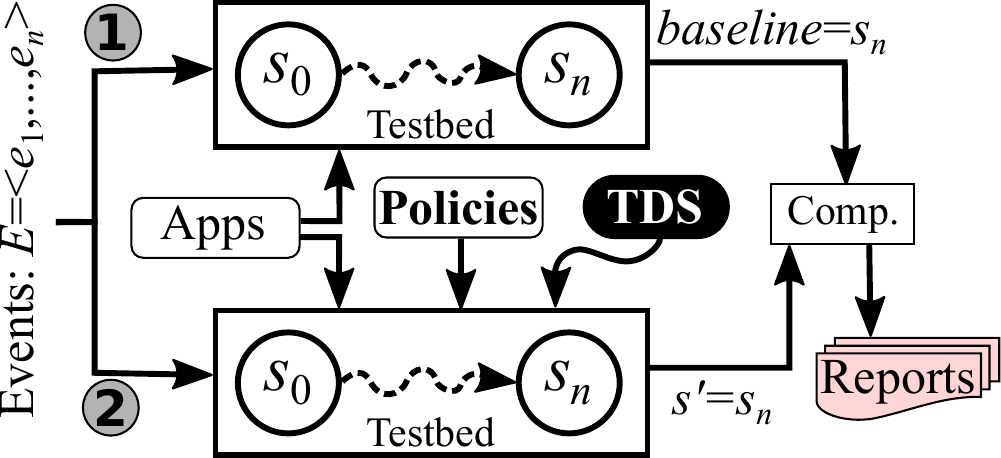}
	\caption{Testing a target defense solution (TDS) using \sysname}
	\label{fig:oracleSPrimeComaparision}
\end{figure}

To overcome this challenge, we adopted a heuristic approach to externally measure the efficacy of a TDS
(see \FIGREF~\ref{fig:oracleSPrimeComaparision}).    
Technically, we run each testcase twice: once in the vanilla testbed $\TB_{\Oracle}$ (path \circledOne) 
and once in the TDS's testbed $\TB_\TDS$ (path \circledTwo). Recall that both $\TB_{\Oracle}$ 
and $\TB_\TDS$ both have the same set of devices and apps, but only $\TB_\TDS$ contains the policies
and the solution code $\TDS$. The final system-state $s_n$ of $\TB_{\Oracle}$ serves as the ${baseline}$,
which describes what the final system-state would be without the defense solution \TDS.
The $baseline$ will be compared against the final system-state $s_n$ of $\TB_\TDS$ (\ie, $s'$).

After collecting both $baseline$ and $s'$, \sysname uses the comparator (see Algorithm~\ref{algo:sysytemStateComparison})
which operates as follows:  
for each device $i$, it checks if the observed state of the device ($d_i$) in $baseline$ is different  
from that in $s'$. Any discrepancy indicates that some actions were allowed in the vanilla testbed $\TB_\Oracle$
but later blocked in the \TDS's testbed $\TB_\TDS$, indicating
the possibility of \TDS becoming successful in preventing unexpected actions.
The comparator next checks if both $s_0$ and $s'$ have recorded the same state
of the device. The equality here means that  the initial state of the device did not change in the final system-state, 
implying the success \TDS in preventing unexpected actions on this device. 
However, if $d_i$ of $s_0$ and $d_i$ of $s'$ are unequal, then the comparator cannot be conclusive
in deciding about the success of \TDS. Therefore, the comparator marks them as ``{indeterminate}''. 

\begin{figure}
	\centering
	\begin{minipage}{0.85\linewidth}

\begin{algorithm}[H]
\caption{Comparator}
\label{algo:sysytemStateComparison}
\begin{algorithmic}[1]
\REQUIRE Initial system-state $s_0=\langle d_1, d_2, .... d_m\rangle,$ \\ $baseline=\langle d_1, d_2, .... d_m \rangle,$
 $s'=\langle d_1, d_2, .... d_m \rangle$, assuming there exists $m$ devices
	\STATE \code{policyViolation} $\gets$ \code{false}
	\STATE \code{indeterminate}  $\gets$ \code{false}
	\FOR {$i \gets 1$ to $m$}
		\IF{ $baseline[d_i] \ne s'[d_i]$}
			\STATE \algorithmiccomment{Defense might work}
			\IF{$\mathit{s'[d_i] = s_0[d_i]}$} 
\STATE \code{policyViolation} $\gets$ \code{true} \algorithmiccomment{Defense worked}
			\ELSE
\STATE \code{indeterminate}  $\gets$ \code{true} \algorithmiccomment{Unsure}
			\ENDIF
		\ENDIF
	\ENDFOR 
	\IF {\code{indeterminate}  = \code{true}}
		\STATE Report ``the outcome is \code{indeterminate}''
	\ELSIF{\code{policyViolation} = \code{true}}
		\STATE Report ``the TDS prevented unexpected actions''.
	\ELSE
		\STATE Nothing to report
	\ENDIF
\end{algorithmic}
\end{algorithm} 	\end{minipage}
\end{figure}

To help an evaluator debug such indeterminate cases, \sysname offers a debug mode,
which is slow but generates a detail report to provide fine-grained insights of the testbed
for the testcase in question. When the debug mode is enabled,
the comparator collects the system-state of the testbed after pushing each event of the sequence, 
unlike the normal mode where it waits for the entire sequence to finish running.
After running \sysname in the debug mode for the testcase twice: once in $\TB_{\Oracle}$ and later in $\TB_{\TDS}$,
the comparator generates a trace file containing the pairwise differences 
between the system-state of $\TB_{\Oracle}$
and that of $\TB_{\TDS}$ at every step. The evaluator can utilize this detail report
to manually inspect such cases.

\subsection{Automated Event Generators}
\label{sec:event-gens}

\sysname uses an event sequence generator to provide the sequence of events for each test case. 
If not configured to test a predefined set of event sequences, 
\sysname defaults to one of two automated event generators: \textit{random} or \textit{realistic}.

\noindent
\textbf{Random Event Generator.} Researchers commonly evaluate smart home defense systems using sequences of randomly generated events 
\cite{contexiot17,mazhar2023maverick,iotguard2019ndss}. 
To facilitate this, \sysname includes a random event sequence generator. 
For each test case, the generator produces an event sequence 
$E_i = \langle e_1, e_2,\ldots \rangle$, where $E_i$ represents the event sequence for the $i$-th testcase $\TSc_i$. 
The sequence length $|E_i|$ is randomly selected between one and the maximum length defined in the configuration. 
To avoid generating irrelevant events, the generator takes into account \Events (as discussed in $\S$~\ref{sec:model}), 
which is based on the installed devices in the testbed. It then randomly selects a device and generates an event 
$e_j$, which could either be an activity supported by the device or a status update reflecting a change in the device's state.

Although simple to implement, this generator has a drawback. While it ensures that each $e_j$ is a plausible event for the testbed, the event order is entirely random, unaffected by automation apps or typical user activities in a smart home. As a result, many sequences are unlikely to occur in real-world scenarios or trigger automation apps, reducing testing efficiency in terms of time and cost. Therefore, in addition to random event sequences, it is crucial to evaluate defense systems using realistic sequences that reflect the influence of automation apps in a typical smart home.

\noindent
\textbf{Realistic Event Generator.}
To enable testing with realistic sequences, \sysname leverages a language model proposed by \helion \cite{manandhar2020Helion}. 
This model generates smart-home events influenced by user-driven automation apps and typical household activities. 
Technically, it predicts the next likely event based on the context, \ie, the history of past events. 
\sysname constructs realistic event sequences by combining the events predicted by the model and 
uses these sequences as test cases to evaluate defense systems, offering an alternative to random event sequences.

\helion developed an $n$-gram statistical language model using a dataset of 273 smart-home apps 
collected through a user survey of 40 participants. 
The model is based on the principle that smart-home apps, created by humans to automate particular workflows, exhibit regular patterns
like natural languages.
\helion trained several $n$-gram models to learn the conditional probability of an event occurring after 
a sequence of prior events (\ie, the context). 
The model can predict either the most probable event (with the highest conditional probability) or the least probable event (with the lowest conditional probability) based on the context. 
Rather than considering the entire event history, \helion found that models using $n=3$ or $n=4$ offered the best 
predictive accuracy, as longer sequences led the model to learn less common patterns.

\helion's statistical models were trained on its own dataset (\ie, user-driven automation apps collected from its survey). 
Consequently, some events predicted by the model may not align with the specific testbed used by \sysname for evaluating IoT defenses. 
To address this, \sysname incorporates a filtering mechanism that wraps around \helion's model. 
This mechanism discards any events unsuitable for the current testbed by referencing the testbed configuration.

\section{Implementations}
\label{sec:impl}
We implemented a fully functional prototype of \sysname 
using Python 3.9. As a the programmable IoT platform, 
\sysname utilizes the \oh platform \cite{OpenHab} (precisely, \oh-3.2.0 stable runtime edition).
During an evaluation of a target defense solution (TDS), different components
of \sysname communicates with the platform using \oh's REST API.  
\sysname accepts testbed configurations in toml format.
It uses json format for internal usage and reporting results.
The event simulator uses Python's random module for randomly generating
event sequences. 
\sysname introduces some delays at several points during an evaluation.
For example, the mechanism to install apps in \oh does not instantaneously
load freshly installed apps. Therefore, \sysname waits for 10 seconds (resp., 15 seconds)
when it installs vanilla apps (resp., instrumented apps) in the testbed.
To ensure that each injected event is processed by the platform and that the triggered apps
have enough time to finish their execution, \sysname waits for 10 seconds. 
All the delay times were selected based on our experiments with the platform where 
we gradually increased the delay times until we found a suitable duration when 
\sysname could finish fidelity testing without any incomplete execution of the apps.

For the \helion-based event generator, we first trained the $n$-gram language models offline 
using \helion's publicly available implementation.\footnote{\url{https://github.com/Secure-Platforms-Lab-W-M/Helion-on-Home-Assistant}}
We then implemented this generator to invoke the specified model during evaluation as requested by the configuration. 
Since \sysname is built on the \oh platform, and \helion's implementation does not use \oh-specific terminology 
for devices and events, we created an on-the-fly conversion process using a \helion-to-\oh mapper. 
One author manually created the mapping between the two terminologies, while another independently verified it to ensure accuracy.

Several defenses like \iotguard and \maverick employ multiple components beyond the IoT platform.
To achieve better isolation, reproducibility, and efficient deployment, we enhanced \sysname 
by adding support for Docker containers to containerize those components. 
Additionally, to conduct multiple experiment sessions in parallel, we automated
the process of instantiating an experiment with \sysname inside a virtual machine by using 
Vagrant.

 \section{Evaluation}
\label{sec:eval}

We now demonstrate how our prototype implementation of \sysname
empirically assessed several IoT defenses that enforce policies at runtime. For our evaluation, 
we selected four test-subjects: \expat \cite{yahyazadeh2019expat}, \patriot \cite{yahyazadeh2020patriot}, 
\iotguard \cite{iotguard2019ndss}, and \maverick \cite{mazhar2023maverick}. 
While they are highly popular in the community and 
closely related in terms of their policy enforcement
mechanism, none of them was empirically compared with each other. 
Another goal of this evaluation is to show how the developers
of those defense solutions could have utilized \sysname had
it existed back then.

The research question we set out to answer is: \textit{can \sysname automatically
evaluate each of these four IoT defenses and compare them empirically?} 
To answer this question, we conducted the evaluation using 
three testing approaches -- 
fidelity, stress and differential -- discussed in $\S$~\ref{sec:prob-def}.
As our evaluation metrics, we used two counts: 
``\code{violation}'' and ``\code{indeterminate}''. 
While the former denotes 
the number of testcases for which \sysname detects that the defense solution 
managed to prevent unexpected actions, the latter 
denotes the number of testcases for which \sysname is unsure
whether the defense prevented unexpected actions or not. The remaining testcases are
all considered as ``\code{no violation}''.
We also measured the time and space (memory) required
by \sysname.

\subsection{Setup}

In our evaluation, 
we used the open-source implementations
of \expat, \patriot,  and \maverick from their respective public repository (\ie, github.com).
Despite \iotguard's popularity, unfortunately, a full implementation of \iotguard
is not publicly available. Therefore, we re-implemented \iotguard by closely following \cite{iotguard2019ndss}. 
Although \iotguard was originally implemented for \smt, we chose the \oh platform to implement \iotguard
as the remaining IoT defenses (\expat, \patriot, and \maverick) were implemented for \oh.
While both \expat and \patriot implement all their policy enforcement modules (\ie, PIP, PEP, PDF)
in the platform, \iotguard employs a remote offshore server to host its PDF and exports
data from the platform to the server for policy decision. Therefore, our in-house \iotguard implementation
spawns a separate server to host its PDF and intermediate data necessary for making policy decision
(for details, see Appendix~\ref{sec:appen:iotguard}).
On the other hand, \maverick implements its all modules (\ie PIP, PEP, and PDF) on a trusted intermediary node (aka, a proxy)
that intercepts and monitors communications 
between IoT devices and the platform.
For isolation, reproducibility, and efficient deployment, we containerized components of IoT defenses that
operate outside the platform. 
For example, the policy server of \iotguard and the proxy of \maverick are deployed in 
individual docker containers.

While \sysname can utilize a testbed with physical devices for experiments,
we chose virtual devices for our evaluations, 
because they were used by our test-subjects in their original evaluation.
For fidelity and stress testing, we replicated the same virtual smart home and policies
as used in their respective paper (recall $\TB_{\expat}, \TB_{\patriot}, \TB_{\iotguard},   \TB_{\maverick}$ 
in $\S$~\ref{sec:model}). 
However, for differential testing, we used an identical smart home (\iotsystem) and the same set of policies (\Policies)
for all test subjects. 
Differential testing were conducted on two pairs of (\iotsystem, \Policies). 
The first pair was chosen from \expat \cite{yahyazadeh2019expat}, the lowest common denominator, 
whereas the second pair was based on \helion \cite{manandhar2020Helion}, a research dataset
aiming to generate natural smart home scenarios.
While \expat's \iotsystem had only one set of IoT applications, \helion's \iotsystem contains 40 sets IoT applications.
Instead of using all 40 IoT application sets,
we chose the top 5 sets with the highest number of applications and then created 5
virtual smart homes using them (\eg, $\iotsystem_{Helion\_1}, \ldots, \iotsystem_{Helion\_5}$). For each smart home ($\iotsystem_{Helion\_x}$) based on \helion, we used the same set of policies 
$\Policies_{Helion}$ mentioned in \helion.
In total, for differential testing, 
we evaluated all 4 IoT defenses on 6 identical setups: 1 based on \expat and 5 based on \helion.

The event sequence (\EventSeq) used in a testcase depends on the testing approach. 
For fidelity testing, we hand-crafted each event sequence for \sysname 
according to the testcases used
in the test-subject's original evaluation. 
For stress testing and differential testing, we utilized both automated event generators
of \sysname (\ie, the random event generator and the \helion-based event generator in $\S$~\ref{subsec:SystemWorkflow}). 
For stress testing, 
we evaluated each IoT defense twice: once with randomly generated events and next with \helion generated events. 
The set of testcases used in stress testing is unique for each test subject. 
On the contrary, for differential testing, all test subjects were tested against the same
testcases. Recall that we used 6 identical testbed setups for differential testing (1 
based on \expat and 5 based on \helion). 
While conducting differential testing on each IoT defense 
using \expat's setup ($\iotsystem_{\expat}, \Policies_{\expat}$),
we reused the same set of testcases that we generated using the random event generator. Contrarily, for differential testing using \helion's testbeds 
($\iotsystem_{Helion\_x}, \Policies_{Helion}$),
we reused the same set of testcases that we generated using the \helion event generator.

\helion's models can generate two types of automation-influenced events: \textit{up} and \textit{down}. 
The \textit{up} flavor produces highly probable event sequences, 
while the \textit{down} flavor generates the least probable ones. 
In other words, the \textit{up} flavor predicts events that follow regular patterns from the user-driven automation used for training, whereas the \textit{down} flavor predicts events that deviate from these patterns. 
Testing with \textit{up} sequences is suitable for evaluating the correctness of a smart home's operational behavior 
and addressing benign issues like functionality errors, inconsistent outcomes, and interference. 
In contrast, \textit{down} sequences are used to assess a system's resilience against attacks \cite{manandhar2020Helion}. 
Static analysis-based defense approaches, such as Soteria \cite{celik2018soteria} and IoTSan \cite{nguyen2018iotsan}, 
are designed to detect and resolve issues caused by regular events before deployment. 
On the other hand, dynamic policy enforcement approaches, like \iotguard \cite{iotguard2019ndss} and \expat \cite{yahyazadeh2019expat},
aim to protect smart homes against unexpected or irregular events (or attacks) at runtime. 
Since \sysname is specifically designed to test dynamic policy-enforcing IoT defenses, we used 
\textit{down} flavor events in our experiments.

We configured \sysname to create 6 test-suites consisting of 5, 10, 15, 25, 35, and 50
different testcases. Each testcase can have at most 15 events, while the actual number of events was
varied between 1 and 15.
All our experiments were conducted in a server machine equipped with 128 core 2.0 GHz 
Intel(R) Xeon(R) Gold 6338 CPU and 256GB RAM.
For each experiment, \sysname created a virtual machine equipped with 
4 cores and 4GB RAM running Ubuntu 20.04 as its operating system.

\subsection{Results}
\label{subsec:results}

\mypara{Fidelity testing.}
With this experiment, we wanted to assess the correctness of our \sysname prototype
by checking whether \sysname could faithfully reproduce the evaluation results of each test-subject
reported in \cite{yahyazadeh2019expat,yahyazadeh2020patriot,iotguard2019ndss, mazhar2023maverick}.
These works conducted their evaluations manually where the authors interacted with the virtual 
devices using the platform UI to create a triggering event and inspected the outcome of each testcase.
\sysname replaced all the manual interaction and inspection with automated approaches while
keeping the same smart home and policies as the prior work.
In this evaluation, we observed that \sysname tested \expat, \patriot, and \maverick with all the testcases
(8, 5, and 8 respectively) and were able to reproduce all results mentioned in 
\cite{yahyazadeh2019expat,yahyazadeh2020patriot,mazhar2023maverick}. 
On the other hand, \sysname reproduced the results of 4 out of 6 testcases for \iotguard. 
Unfortunately, we could not even reproduce the results of the remaining 2 testcases manually 
using the platform's UI. After close inspection, we found that those testcases could not be reproduced solely 
based on the description provided in \cite{iotguard2019ndss}.

\mypara{Stress testing.} Recall that in stress testing, we wanted to evaluate
the efficacy of a given test-subject using \sysname. For each subject, 
we replicated the testbed as mentioned in their original evaluation
but tested the subject with automatically generated test-cases.
We tested each test-subject twice: 
once with the randomly generated 6 test-suites and
next with the \helion generated 6 test-suites.
Technically, we generated a
sequence of events for each testcase required for our test-suites (mentioned earlier). 
In other words, each subject was tested in its original testbed but with different
sets of test-suites. 

\begin{figure*}[!t]
	\centering
	\begin{subfigure}{0.96\textwidth}
\includegraphics[width=\textwidth]{\FIGLOC/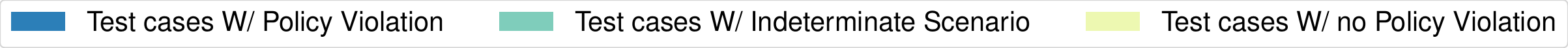}
	\end{subfigure}
	\begin{subfigure}{0.48\textwidth}
		\centering
\includegraphics[width=0.9\linewidth]{\FIGLOC/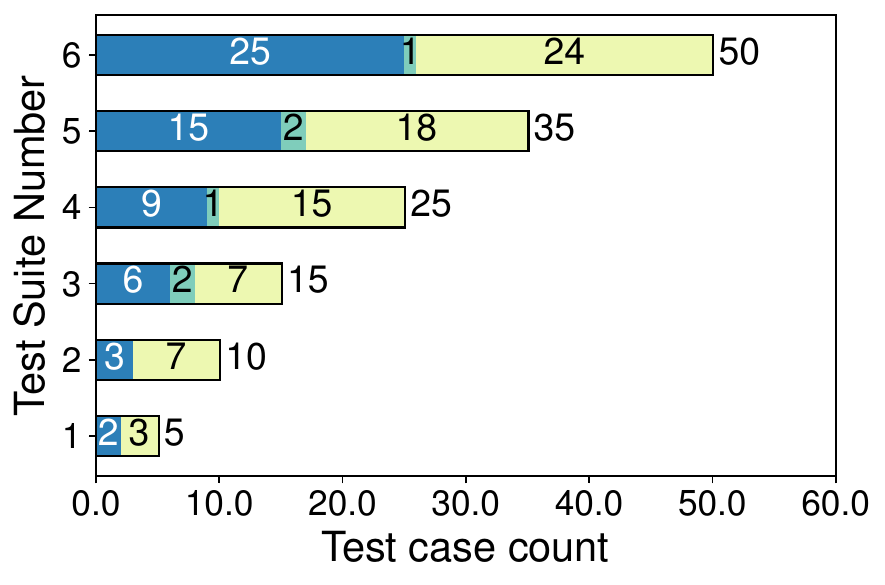}
\caption{\expat}
		\label{fig:ST-expat}
	\end{subfigure}
	\begin{subfigure}{0.48\textwidth}
		\centering
\includegraphics[width=0.9\linewidth]{\FIGLOC/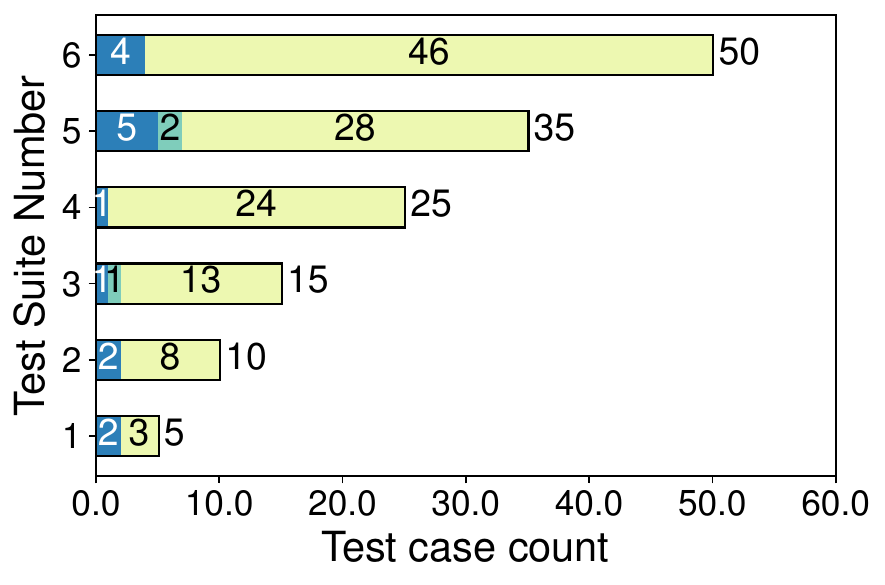}
\caption{\patriot}
		\label{fig:ST-patriot}
	\end{subfigure}
	\begin{subfigure}{0.48\textwidth}
		\centering
\includegraphics[width=0.9\linewidth]{\FIGLOC/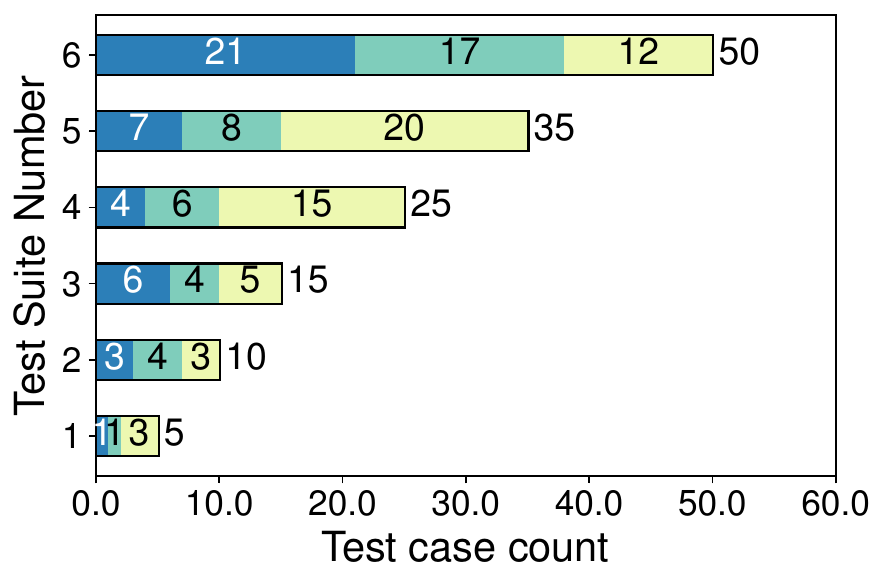}		
\caption{\iotguard}
		\label{fig:ST-iotguard}
	\end{subfigure}
	\begin{subfigure}{0.48\textwidth}
		\centering
\includegraphics[width=0.9\linewidth]{\FIGLOC/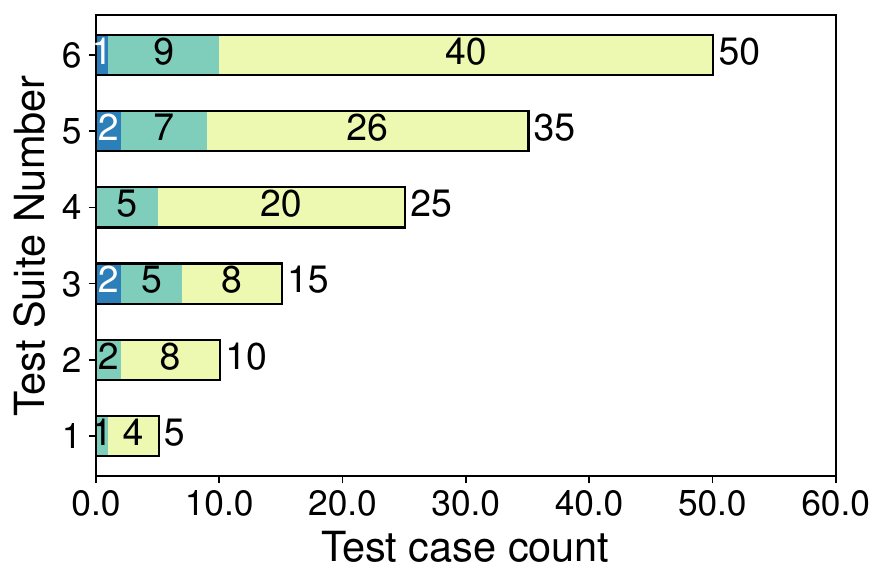}		
\caption{\maverick}
		\label{fig:ST-maverick}
	\end{subfigure}
\caption{Result of Stress Testing IoT defenses with randomly generated events}
	\label{fig:ST}
\end{figure*}

\FIGREF~\ref{fig:ST-expat} shows the outcome of stress testing \expat against 6 randomly generated test-suites.
For example, we observed that \expat was able to prevent unexpected actions
in 25 out of 50 testcases of the test-suite number 6 (\ie, testcases with policy violation).
24 testcases did not cause any policy violation. In the remaining 1 testcase, 
\sysname noticed some changes in one or more devices but was unsure
whether \expat was successful or not, and thus \sysname reported them
as indeterminate.

\FIGREF~\ref{fig:ST-patriot} shows the results of stress testing \patriot against 6
generated randomly test-suites. We observed that \patriot has less number of policy violation and indeterminate cases,
which can be attributed to 
the policy language of \patriot that allows fine-grained policies.
We will explain this benefit with a case study later.

\FIGREF~\ref{fig:ST-iotguard} shows the results of stress testing \iotguard with randomly generated events.
We observed a higher number of policy violation and indeterminate cases, 
which is due to the testbed of \iotguard.
The testbed has many apps that create a loop in multiple testcases 
(\eg, app1's action triggers app2, then app2's action triggers app1).
In such cases, \sysname cannot correctly generate a baseline that is required 
by the comparator algorithm.

\FIGREF~\ref{fig:ST-maverick} shows the results of stress testing \maverick with randomly generated events.
We observer that the number of indeterminate cases was higher than the number of policy violation.
This can be attributed to \maverick's recovery mechanism, which kicks in 
when \maverick could not prevent an unexpected physical action on a device 
and applies corrective actions as a remedy.

\begin{figure*}[!t]
	\centering
		\begin{subfigure}{0.96\textwidth}
\includegraphics[width=\textwidth]{\FIGLOC/bar-plt-legend-helvetica-20.pdf}
		\end{subfigure}
	\begin{subfigure}{.48\textwidth}
		\centering
\includegraphics[width=0.9\linewidth]{\FIGLOC/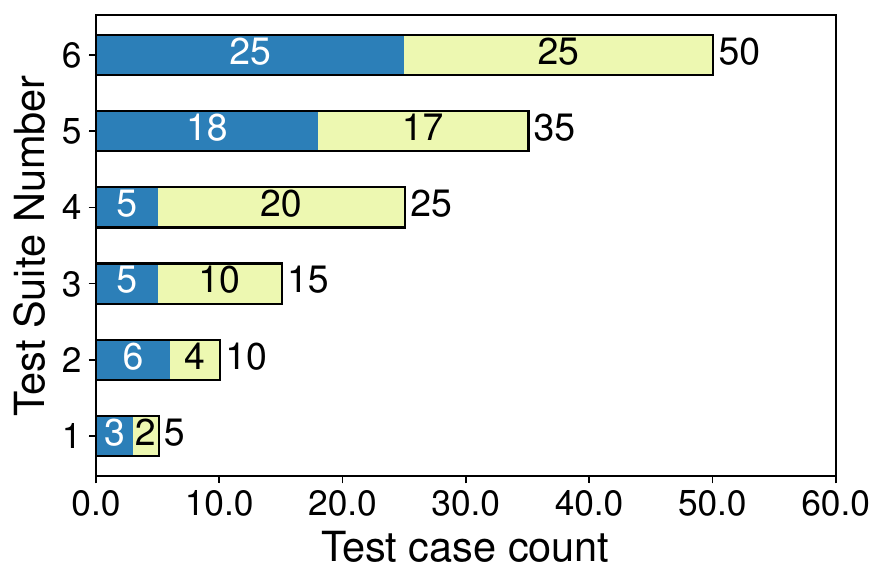}
\caption{\expat}
		\label{fig:ST-expat-with-helion}
	\end{subfigure}
	\begin{subfigure}{.48\textwidth}
		\centering
\includegraphics[width=0.9\linewidth]{\FIGLOC/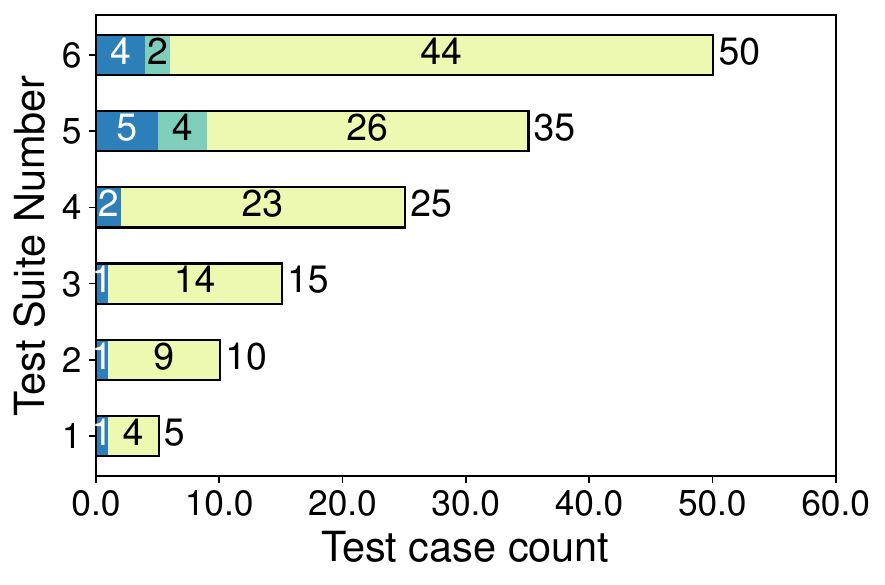}
\caption{\patriot}
		\label{fig:ST-patriot-with-helion}
	\end{subfigure}
	\begin{subfigure}{.48\textwidth}
		\centering
\includegraphics[width=0.9\linewidth]{\FIGLOC/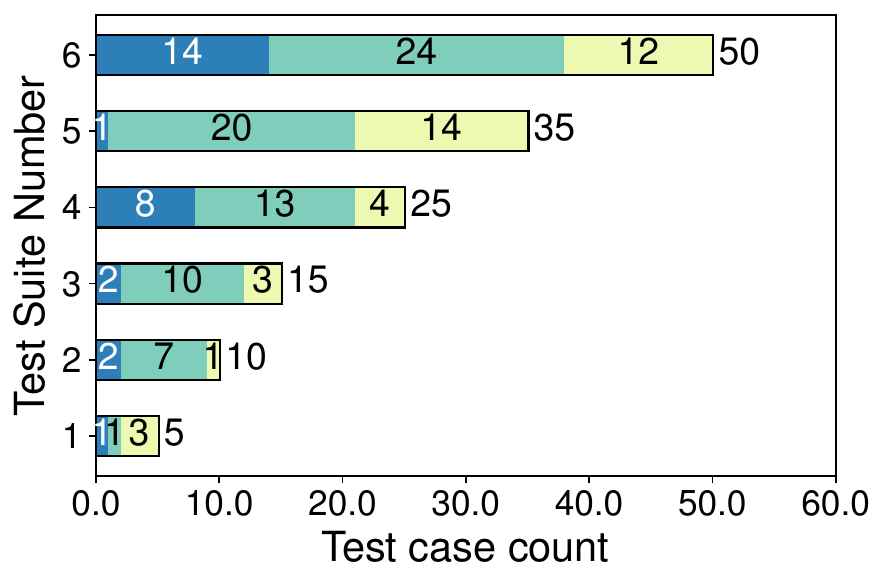}		
\caption{\iotguard}
		\label{fig:ST-iotguard-with-helion}
	\end{subfigure}
	\begin{subfigure}{.48\textwidth}
		\centering
\includegraphics[width=0.9\linewidth]{\FIGLOC/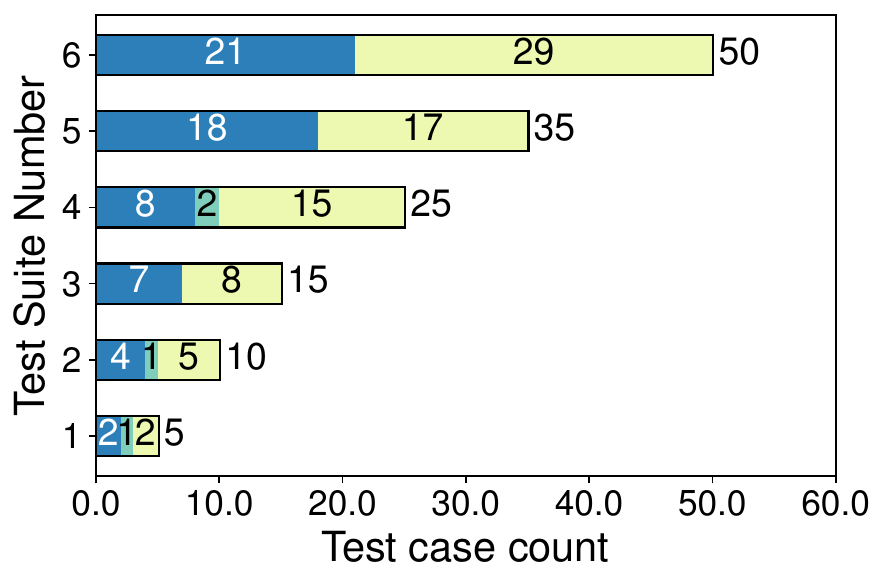}		
\caption{\maverick}
		\label{fig:ST-maverick-with-helion}
	\end{subfigure}
\caption{Result of Stress Testing IoT defenses with \helion-based generated events}
	\label{fig:ST-with-helion}
\end{figure*}

\FIGREF~\ref{fig:ST-with-helion} shows the result of stress testing all 4 IoT defenses
in their original testbeds but
against the test-suites created using the \helion-based event generator.
\FIGREF~\ref{fig:ST-expat-with-helion} shows the outcome of stress testing \expat 
against the six \helion generated test-suites.
Compared to \FIGREF~\ref{fig:ST-expat}, the \helion based event generation results in 
higher policy violation count for test-suite 1, 2 and 5.
Similarly, 
\FIGREF~\ref{fig:ST-patriot-with-helion}, \ref{fig:ST-iotguard-with-helion}, and \ref{fig:ST-maverick-with-helion}  
show the outcome of stress testing \patriot, \iotguard, and \maverick, respectively, using 
the six \helion generated test-suites. 
We observed some variations. For instance, the \helion based event generation resulted in more indeterminate cases for \iotguard, compared to 
the \iotguard's outcome when tested against randomly generated events.

Overall, when compared to \FIGREF~\ref{fig:ST}, \helion generated test-suites (shown in \FIGREF~\ref{fig:ST-with-helion})
resulted in slightly more policy violation or indeterminate cases. 
This can be attributed to the naturalness of \helion-based test scenarios, as 
events are generated based on a statistical model trained with diverse 
user-driven automation workflows.
This ensures that the \helion-based events include realistic, but irregular, activities for typical smart homes. 
Compared to randomly generated events, these realistic events lead to more executions of IoT apps, 
resulting in a frequent checking of installed policies.

\mypara{Differential testing.}
In this experiment, we wanted to assess \sysname's 
capability to empirically compare the test-subjects. 
\sysname used identical testbeds along with 
the same policies (written in the subject's language)
and the same suites of testcases to evaluate each subject.
Recall that we used 6 identical testbeds for differential testing. 
We compared the test outcomes of each subject with that of others.
The rational was to measure how a subject fared compared to others in preventing
unexpected actions.

\begin{figure*}[!t]
	\begin{subfigure}{0.96\textwidth}
\includegraphics[width=\textwidth]{\FIGLOC/bar-plt-legend-helvetica-20.pdf}
	\end{subfigure}
	\begin{subfigure}{0.49\textwidth}
		\centering
\includegraphics[width=1\linewidth]{\FIGLOC/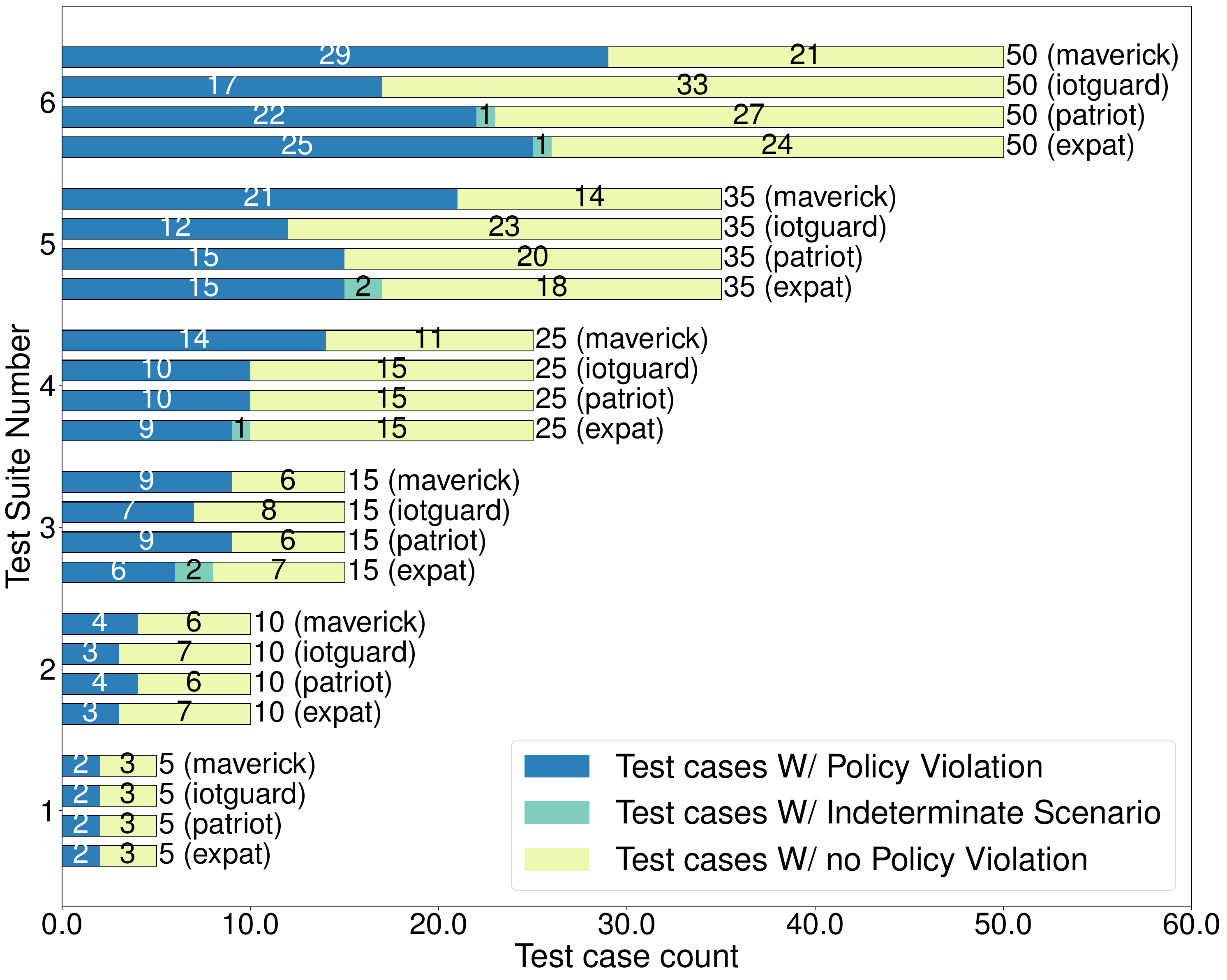}
\caption{ Testbed: \expat}
		\label{fig:DT-expat}
\end{subfigure}
	\begin{subfigure}{0.49\textwidth}
		\centering
\includegraphics[width=1\linewidth]{\FIGLOC/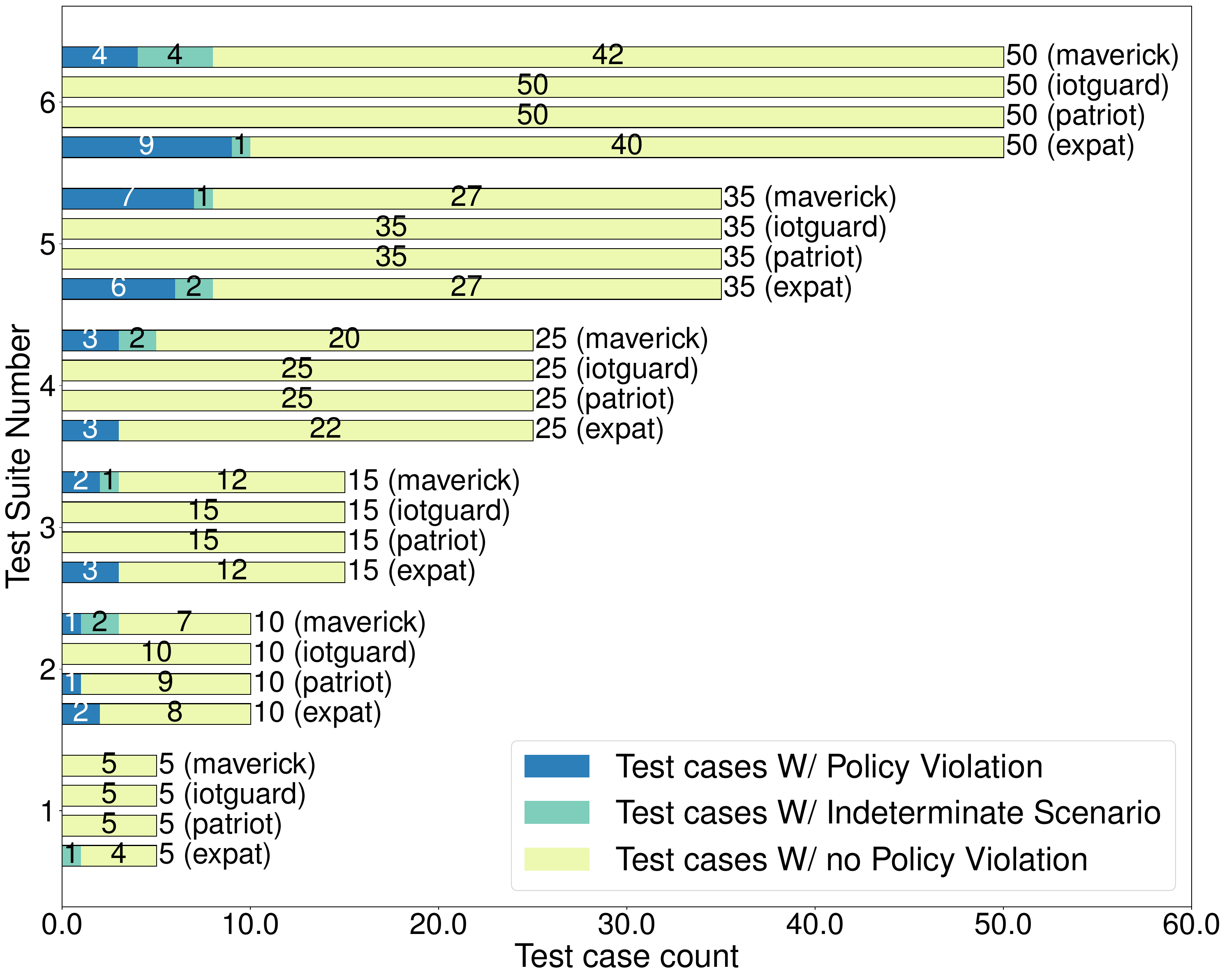}
\caption{ Testbed: \helion-1}
		\label{fig:DT-helion-1}
\end{subfigure}
\caption{Result of Differential Testing with \sysname (Group 1 of 3)}
\end{figure*}

\FIGREF~\ref{fig:DT-expat} shows the comparison of differential testing using \expat's testbed ($\iotsystem_{\expat}, \Policies_{\expat}$). 
We observed that
\maverick detected higher number of policy violations than that of others.
\expat and \patriot behaved similarly in most cases. \iotguard was able to prevent relatively lower number of policy violation.  The outcome of \maverick can be attributed to the fact that \maverick performs policy checks 
whenever it intercepts an event or action, irrespective of whether the event or action 
triggers the execution of an IoT application. In contrast, other defenses only check 
for policy violations when an action is issued by an IoT application. Some of our test cases 
did not trigger the execution of any installed IoT applications, resulting in fewer reported policy violations.
The outcome of \iotguard can be attributed to its policy enforcement mechanism, which
will be further explained in a case study later ($\S$~\ref{subsec:case-analysis}).

\begin{figure*}[!t]
	\begin{subfigure}{0.96\textwidth}
\includegraphics[width=\textwidth]{\FIGLOC/bar-plt-legend-helvetica-20.pdf}
	\end{subfigure}
	\begin{subfigure}{0.49\textwidth}
		\centering
\includegraphics[width=1\linewidth]{\FIGLOC/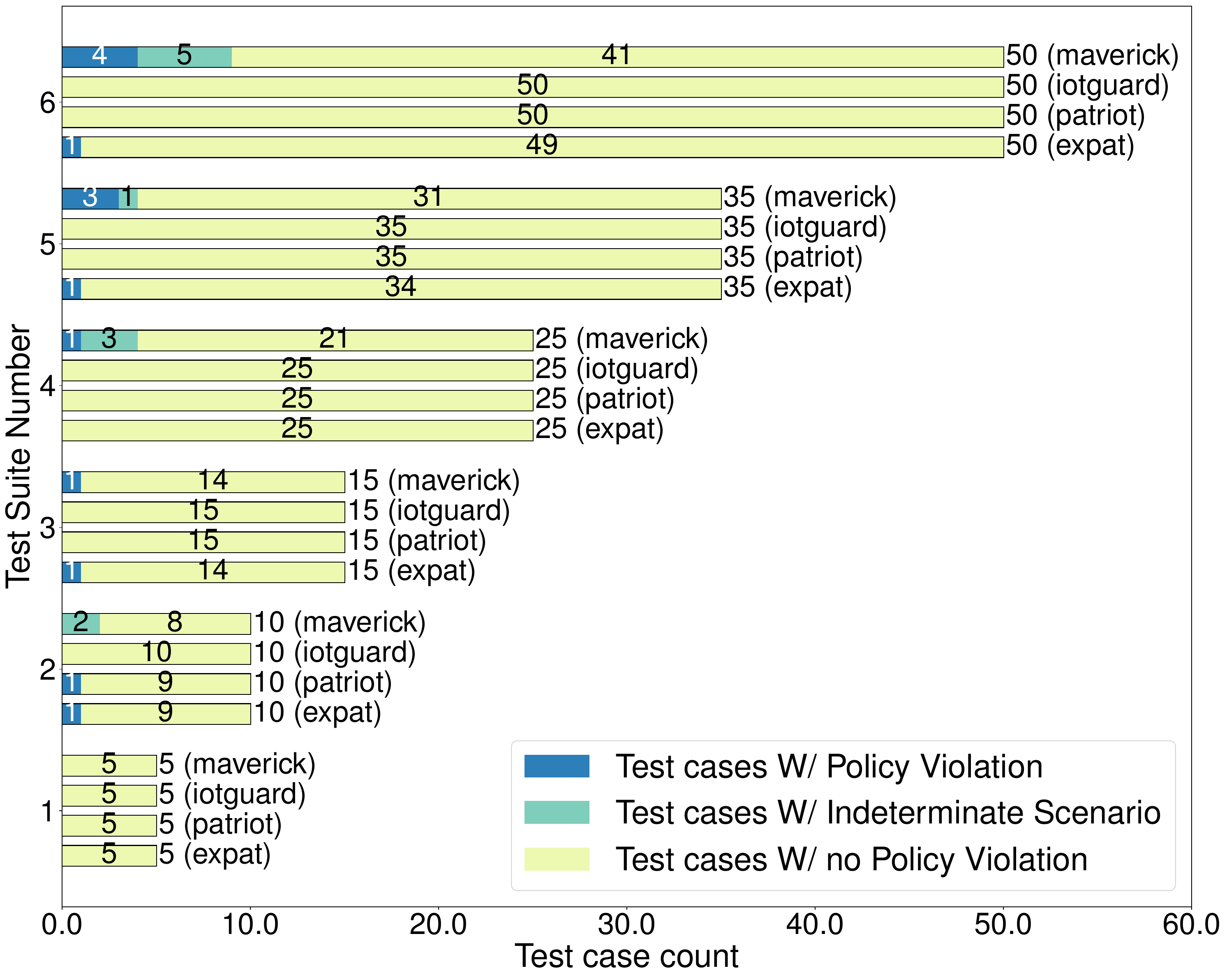}
\caption{ Testbed: \helion-2}
		\label{fig:DT-helion-2}
\end{subfigure}
	\begin{subfigure}{0.49\textwidth}
		\centering
\includegraphics[width=1\linewidth]{\FIGLOC/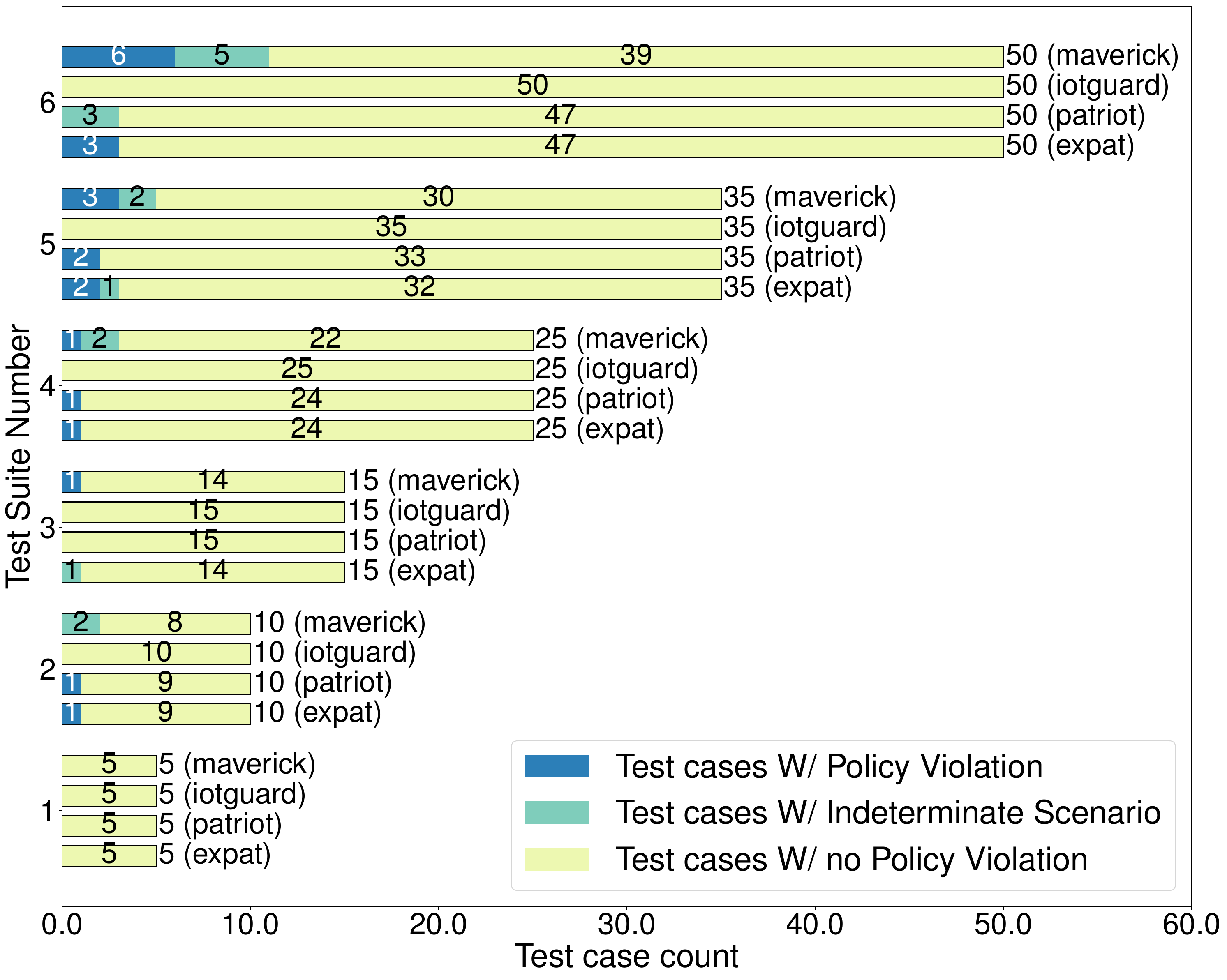}
\caption{ Testbed: \helion-3}
		\label{fig:DT-helion-3}
\end{subfigure}
\caption{Result of Differential Testing with \sysname (Group 2 of 3)}
\end{figure*}

\FIGREF~\ref{fig:DT-helion-1}, \ref{fig:DT-helion-2}, \ref{fig:DT-helion-3}, 
\ref{fig:DT-helion-4}, and \ref{fig:DT-helion-5} show the comparison
of differential testing using 5 \helion-based testbeds (\ie, $\iotsystem_{Helion\_1}, \ldots, \iotsystem_{Helion\_5}$). 
For each testbed, we used the same \helion-based policies ($\Policies_{Helion}$).
One noticeable pattern is that the policy violation counts (similarly, indeterminate counts)
are much lower than that of \FIGREF~\ref{fig:DT-expat}. 
We attribute this difference to the characteristics of IoT applications 
used in \expat's and \helion-based testbeds.
As per \expat's original evaluation \cite{yahyazadeh2019expat},
the testbed employed a small set of deliberately crafted IoT applications (15 in total) 
designed to violate some of the installed policies ($\Policies_{\expat}$).
Consequently, the execution of an app was more likely to result in a policy violation, 
thus explaining the higher counts of policy violations.
In contrast, the \helion dataset comprises a large pool of IoT applications (273)
collected from 40 smart home users. 
These applications were not designed with policy violation detection in mind, 
reducing the likelihood of a dynamic relationship between an application and policy violations. 
This discrepancy accounts for the lower counts of policy violations observed in the \helion-based testbeds.

\begin{figure*}[!t]
	\begin{subfigure}{0.96\textwidth}
\includegraphics[width=\textwidth]{\FIGLOC/bar-plt-legend-helvetica-20.pdf}
	\end{subfigure}
	\begin{subfigure}{0.49\textwidth}
		\centering
\includegraphics[width=1\linewidth]{\FIGLOC/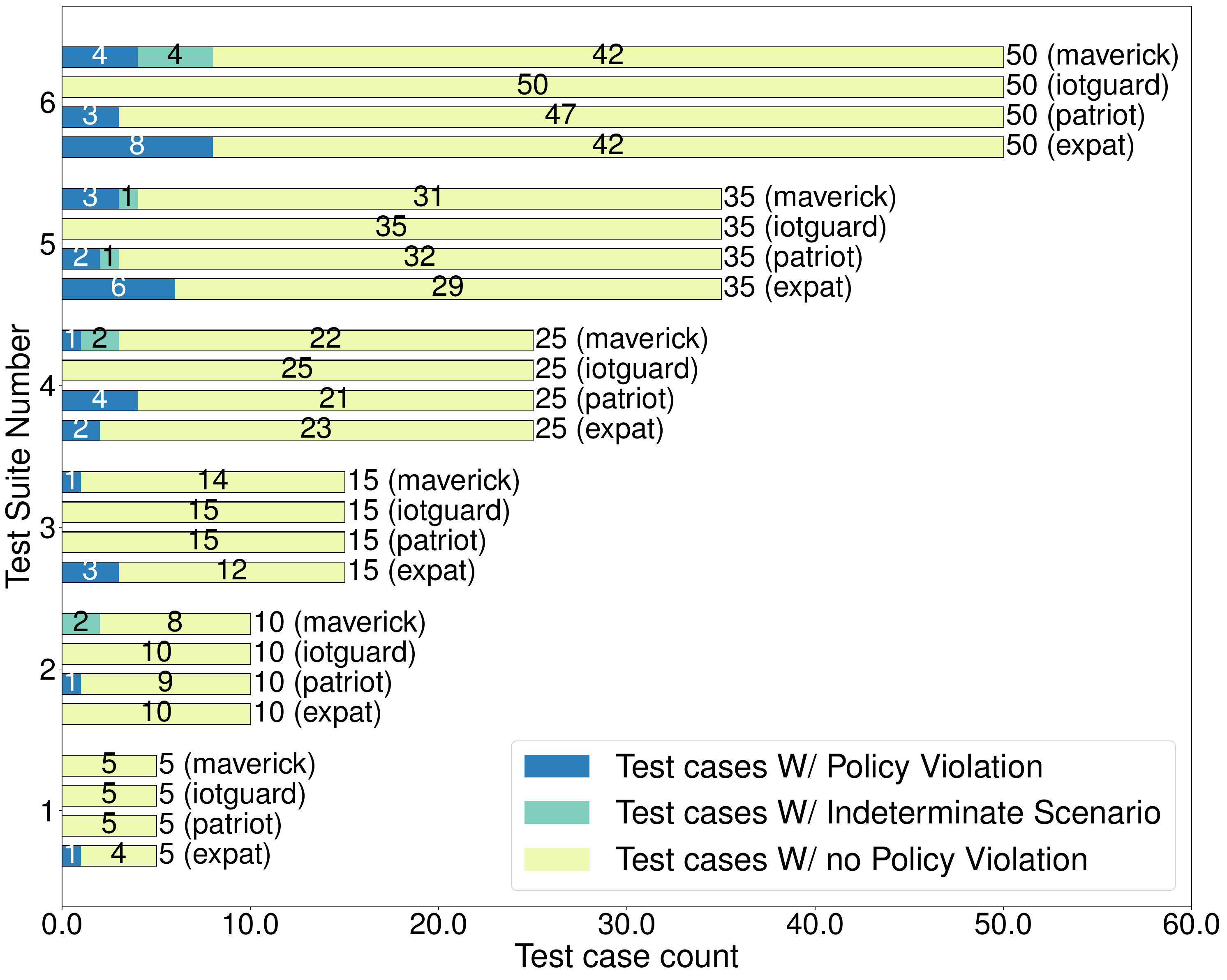}
\caption{ Testbed: \helion-4}
		\label{fig:DT-helion-4}
\end{subfigure}
	\begin{subfigure}{0.49\textwidth}
		\centering
\includegraphics[width=1\linewidth]{\FIGLOC/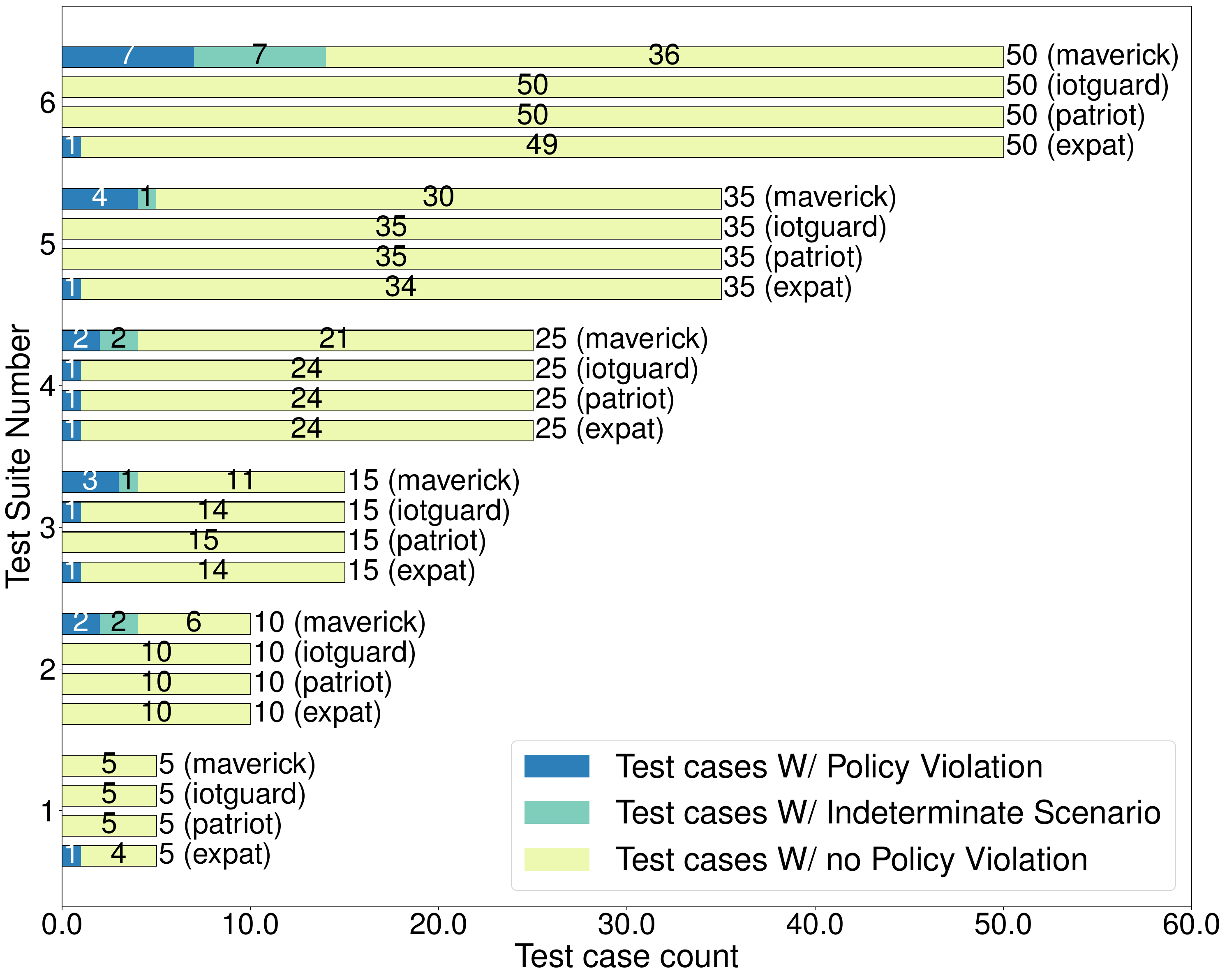}
\caption{ Testbed: \helion-5}
		\label{fig:DT-helion-5}
\end{subfigure}
\caption{Result of Differential Testing with \sysname (Group 3 of 3)}
\end{figure*}

Another pattern we observed is that mostly \maverick and \expat reported policy violations in the 
\helion-based testbeds compared to \iotguard and \patriot.
Since \maverick conducts policy checks whenever it intercepts 
either an action or an event, the likelihood of detecting a policy violation is doubled,
which explains the higher count of policy violations.
In contrast, although \expat only checks 
for policy violations when an action is issued by an IoT application, 
it reports more policy violations compared to \patriot and \iotguard. 
This is because \expat does not employ a policy selection algorithm when it checks policies, unlike \patriot and \iotguard.
As a result, \expat erroneously reports a violation of a policy that is not even relevant
to the intercepted action.
A similar case study is described in $\S$~\ref{subsec:case-analysis} (see Type2).

\mypara{Time and Space Requirement.}
To measure empirical
time and space complexity of \sysname, 
we recorded its peak memory usage and the total required time for each experiment, running all six test-suites.
In total, there were 32 experiments: 8 for stress testing and 24 for differential testing ($= 4$ IoT defenses $\times$ $6$ identical
testbeds).
\FIGREF~\ref{fig:cpu-time} and \ref{fig:mem} show that \sysname required 
less than 5hrs (no more than 60 seconds as CPU-time) 
and less than 2400kB recorded its peak memory usage. Also, the complexity increases linearly with the increase in the number of testcases.
There was an initial high peak memory usage, but after that the memory usage remained nearly constant.
Note that multiple lines overlap with each other on these figures. 

Furthermore,
we categorized our 32 experiments based on each target IoT defense to
clearly present the time and space complexity. We observed that evaluating \maverick required a higher simulation time and consumed
more memory compared to others. 
This can be attributed to \maverick's trusted intermediary node, which 
not only demands additional memory but also increases communication latency. Our observation
is akin to the reported overhead \cite{mazhar2023maverick}.

\begin{figure}[!t]
	\centering
	\begin{subfigure}{0.99\linewidth}
\includegraphics[width=\linewidth]{\FIGLOC/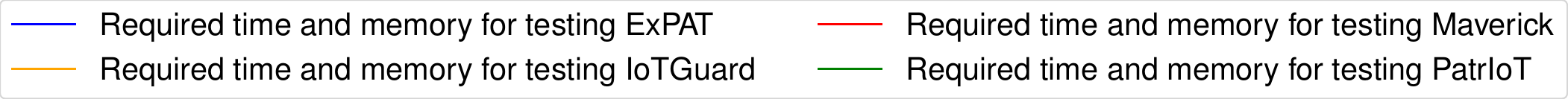}
	\end{subfigure}
	\begin{subfigure}{0.45\linewidth}
		\centering
\includegraphics[width=\linewidth]{\FIGLOC/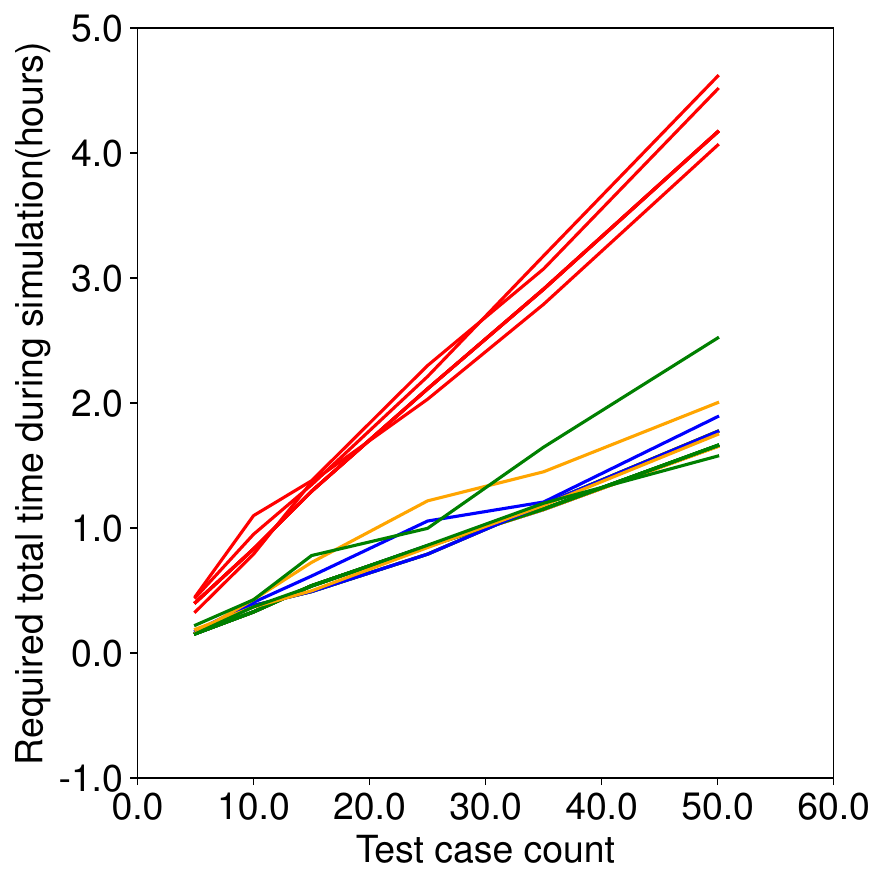}
\caption{Required time}\label{fig:cpu-time}
	\end{subfigure}
	\begin{subfigure}{0.45\linewidth}
		\centering
\includegraphics[width=\linewidth]{\FIGLOC/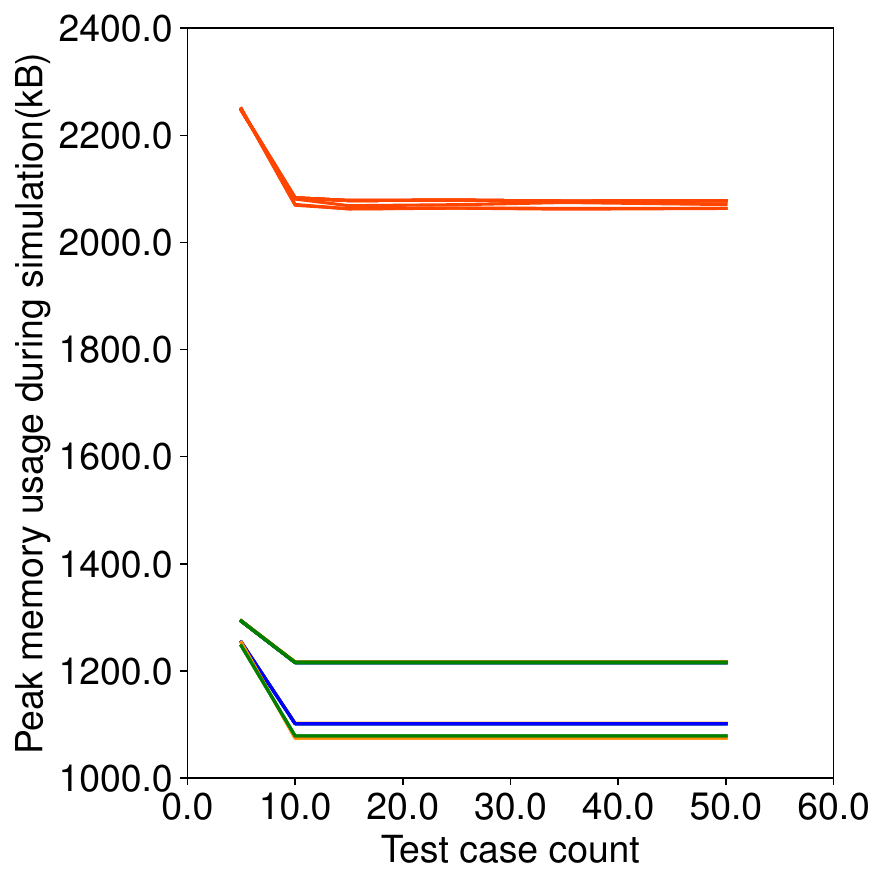}
\caption{Required peak memory} \label{fig:mem}
	\end{subfigure}
\caption{Time and space required by \sysname}
\end{figure}

\subsection{Case Studies}
\label{subsec:case-analysis}

\mypara{Type1: Inconsistency between the \iotguard's policy server and the platform.}
Differential testing revealed that \iotguard failed to prevent policy-violating actions 
in multiple testcases, whereas \expat and \patriot prevented them. 
It is because the \iotguard's policy server and the platform sometimes go out-of-sync, resulting
in erroneous policy enforcement. Note that this issue is specific to \iotguard's design.

Consider a testcase from test-suite 2 of differential testing.
For this testcase,  
\iotguard failed to prevent \code{WaterValve} from turning \code{OFF} when
\code{FireSprinkler} was spraying water due to the smoke detected by \code{SmokeDetector}, even 
with this policy ``\code{PI1: \textit{WaterLeakDetector can turn off WaterValve only if SmokeDetector is OFF}}''
in place.
Upon investigation, we observed that when the platform received the event \code{SmokeDetector=ON}, 
no app was triggered and hence no information about this event was exported to the \iotguard's 
policy server. Recall that \iotguard exports data to the policy server right before an app's 
action block (\ie, PEP). Later when the platform received \code{WaterLeakDetector=ON}, an app
was triggered that contemplated an action \code{WaterValve=OFF}. The PEP of this app 
exported all information 
(\ie, the event \code{WaterLeakDetector=ON} and the contemplated action \code{WaterValve=OFF}).
Since the policy server has no information about \code{SmokeDetector=ON}, 
the server erroneously decided that the contemplated action would not violate \code{PI1}
and thus allowed the action, causing \code{WaterValve} to be \code{OFF} 
at a wrong time.

\mypara{Type2: Efficacy depends on the policy selection process.}
How a defense solution selects which 
policies to evaluate during its policy enforcement impacts its efficacy. 
While \patriot and \iotguard select only those polices that are relevant to the contemplated
actions to evaluate, \expat always evaluates all policies, which sometimes
results in erroneous enforcement.

Consider a part of a long testcase from test-suite 1 of differential testing, 
where the platform receives the event \code{TV=ON} followed by 
\code{AC=ON} and \code{SleepMode=ON} in that order. 
The \code{TV=ON} event triggered an app that opened 
\code{LivingRoomWindow}. Then when \code{AC=ON} (say, due to a physical action) 
was received, the system reached an unsafe state because it violated a policy
``\code{PI8: LivingRoomWindow can be opened only if both the heater and AC are off.}''
None of the defense solutions prevented it because it was due to a user interaction in the physical world.
 
Later when the \code{SleepMode=ON} event triggers an app that contemplated to turn off \code{TV}, 
\expat denied the action as it evaluated all policies, including \code{PI8} that
turned out to be \code{false}. 
On the contrary, \patriot and \iotguard will allow \code{TV.off()} because they 
did not evaluate \code{PI8} as this policy was not related to the contemplated action.

\mypara{Type3: Erroneous policy decision function.}
\iotguard's policy decision function (PDF) employs a reachability analysis 
on the graph of events and actions that the policy server internally builds based on the data
exported by the platform. During stress testing, \iotguard could not prevent unexpected actions in some testcases due to an erroneous reachability analysis. 
Instead of explaining one of those testcases,
we will use a simple scenario for the sake of brevity.  

Consider a testcase where the event \code{TV=ON} is followed by
the event \code{LivingRoomTemp=90\degree}. The first event
triggered an app that opened \code{LivingRoomWindow}. Later 
the second event triggered an app that contemplated \code{AC.on()}
because \code{LivingRoomTemp>70\degree}. Now as per the policy
``\code{PI8: LivingRoomWindow can be opened only if both the heater and AC are off}'', 
an IoT defense mechanism should not turn on \code{AC} because \code{LivingRoomWindow=OPEN} and 
\code{PI8} is internally represented as ``\code{$\neg$(LivingRoomWindow=OPEN) $\vee$ ((Heater=OFF) $\wedge$ (AC=OFF))}''.

Unfortunately, \iotguard could not block \code{AC.on()} 
because of its reachability analysis based policy enforcement mechanism.
Since there was no direct interaction from the first app to the second app (\ie, one app's
action does not trigger the second app), \iotguard's policy server did not have any path
from the \code{LivingRoomWindow=OPEN} node to the \code{AC=ON} node. As a result, 
\iotguard erroneously decided no violation of \code{PI8} and allowed \code{AC.on()}.

\section{Discussions}
\label{sec:discuss}
\sysname depends on the baseline generation mechanism to measure the impact
of the defense solution. However, a baseline cannot be generated with confidence
when two or more apps create an execution loop (\eg, two apps' actions trigger
each other, creating a loop).

Our in-house implementation of \iotguard faithfully follows
the description \cite{iotguard2019ndss} as closely as possible, and we took some rational
design decisions whenever the description was vague. 
Our evaluation results may differ if the original implementation
of \iotguard is used, since the implementation could include 
optimizations that were not mentioned in \cite{iotguard2019ndss}.

Three of the IoT defenses (\expat, \patriot, and \iotguard) can prevent unexpected actions only when issued by apps. 
As a result, user interactions with the physical devices or on 
the platform UI can potentially move the system into an unsafe state, later causing the defense solutions enforce policy incorrectly. Our randomly 
generated testcases often include such events that push the system into an unsafe state, 
resulting in some indeterminate outcomes.

\section{Related Work}
\label{sec:related}

The rapid growth of IoT innovations has resulted in
multiple avenues of smart-home security research.
Examples include IoT security relying on access control mechanism \cite{rahmati2018tyche, lee2017fact, tian2017smartauth, sikder2020kratos, 10.1145/3460120.3484592, goyal2022securing},
static policy enforcement
\cite{celik2018soteria, chi2020crossapp, nguyen2018iotsan, wang2019charting} 
and dynamic policy enforcement
\cite{wang2018fear, iotguard2019ndss, yahyazadeh2019expat, yahyazadeh2020patriot, ding2021iotsafe, mazhar2023maverick}.
On the contrary, \sysname provides an evaluation platform for 
defense solutions enforcing dynamic policies at runtime.

Testbeds are key component for any kind of research experiment.
Creating testbeds to conduct large scale IoT experiments has become another research problem.
Much of the prior research in this direction focuses on creating large scale physical IoT testbeds. OpenTestBed \cite{munoz:hal-02266558}, LinkLab 2.0 \cite{285062}, 
are physical IoT testbeds that can be accessed via web portal or ssh to conduct IoT experiments. Gotham Testbed \cite{10049670} uses emulation to create IoT devices.
These testbeds are used for IoT data generation and IoT network security experiments.
On the other hand, our target is dynamic policy enforcement based defenses that aim 
to curb unexpected actions issued by automation apps, not devices. Therefore, physical testbeds 
and emulated devices are not essential. Hence, \sysname utilizes an existing IoT platform which offers virtual devices.

Uniform evaluation of IoT security mechanisms has been proposed for other types of IoT systems.
BenchIoT \cite{almakhdhub2019benchiot} 
proposed a test suite and evaluation framework for evaluating defense mechanisms created for the micro-controllers used in IoT devices.
SmartAttack \cite{9290797} proposed a uniform adversarial attack model framework to test security solutions that analyze network trace of IoT devices.
Similarly, 
\sysname provides an evaluation framework for policy enforcing IoT defenses. 

Prior research on generating IoT events has been proposed for security evaluation and simulation purposes.
Helion \cite{manandhar2020Helion} proposes an event generation framework based on patterns found in IoT applications, which can be used to synthesize security policies.

Event generation based evaluation approaches have been proposed for systems.
Android Monkey \cite{androidMonkey} is an UI event generator 
built to test Android applications. Approaches like IoTFuzzer \cite{iotfuzzer}, DIANE \cite{redini2021diane} 
generate app-specific events to test smartphone companion apps for IoT.

\section{Conclusion}
\label{sec:conclusion}

We proposed \sysname, a highly automated uniform evaluation platform
for vetting IoT defense solutions that dynamically enforce security and safety policies at runtime 
to prevent unexpected events/actions in a smart home IoT environment. \sysname replaces much of the 
traditional experimentation process with an automated counterpart, including, testbed 
instantiation, testcase generation and execution, output collection and report generation.
For evaluation, we demonstrated \sysname on four existing IoT defense solutions: \expat, \patriot, \iotguard, and \maverick.
Our results demonstrate that \sysname can be effective in assessing IoT defense solutions.
Researchers can leverage it
to fine-tune their new defense solutions and empirically compare with existing solutions.

\section*{ACKNOWLEDGMENT}
This research was 
supported by the National Science Foundation under
grants CNS-2006556 and CNS-2007512.
Any opinions, findings, and conclusions or recommendations expressed in this material are those of the authors
and do not necessarily reflect the views of the National Science Foundation.

\bibliographystyle{ACM-Reference-Format}
\bibliography{ref}


\begin{thebibliography}{32}


\ifx \showCODEN    \undefined \def \showCODEN     #1{\unskip}     \fi
\ifx \showDOI      \undefined \def \showDOI       #1{#1}\fi
\ifx \showISBNx    \undefined \def \showISBNx     #1{\unskip}     \fi
\ifx \showISBNxiii \undefined \def \showISBNxiii  #1{\unskip}     \fi
\ifx \showISSN     \undefined \def \showISSN      #1{\unskip}     \fi
\ifx \showLCCN     \undefined \def \showLCCN      #1{\unskip}     \fi
\ifx \shownote     \undefined \def \shownote      #1{#1}          \fi
\ifx \showarticletitle \undefined \def \showarticletitle #1{#1}   \fi
\ifx \showURL      \undefined \def \showURL       {\relax}        \fi
\providecommand\bibfield[2]{#2}
\providecommand\bibinfo[2]{#2}
\providecommand\natexlab[1]{#1}
\providecommand\showeprint[2][]{arXiv:#2}

\bibitem[iot({[n.\,d.]})]%
        {iotbench}
 \bibinfo{year}{[n.\,d.]}\natexlab{}.
\newblock \bibinfo{title}{IoTBench-test-suite}.
\newblock
  \bibinfo{howpublished}{https://github.com/IoTBench/IoTBench-test-suite}.
\newblock


\bibitem[Almakhdhub et~al\mbox{.}(2019)]%
        {almakhdhub2019benchiot}
\bibfield{author}{\bibinfo{person}{Naif~Saleh Almakhdhub},
  \bibinfo{person}{Abraham~A Clements}, \bibinfo{person}{Mathias Payer}, {and}
  \bibinfo{person}{Saurabh Bagchi}.} \bibinfo{year}{2019}\natexlab{}.
\newblock \showarticletitle{Benchiot: A security benchmark for the internet of
  things}. In \bibinfo{booktitle}{\emph{2019 49th Annual IEEE/IFIP
  International Conference on Dependable Systems and Networks (DSN)}}. IEEE,
  \bibinfo{pages}{234--246}.
\newblock


\bibitem[Celik et~al\mbox{.}(2018)]%
        {celik2018soteria}
\bibfield{author}{\bibinfo{person}{Z~Berkay Celik}, \bibinfo{person}{Patrick
  McDaniel}, {and} \bibinfo{person}{Gang Tan}.}
  \bibinfo{year}{2018}\natexlab{}.
\newblock \showarticletitle{Soteria: Automated {IoT} Safety and Security
  Analysis}. In \bibinfo{booktitle}{\emph{2018 USENIX Annual Technical
  Conference (USENIX ATC 18)}}. \bibinfo{pages}{147--158}.
\newblock


\bibitem[Celik et~al\mbox{.}(2019)]%
        {iotguard2019ndss}
\bibfield{author}{\bibinfo{person}{Z~Berkay Celik}, \bibinfo{person}{Gang Tan},
  {and} \bibinfo{person}{Patrick~D McDaniel}.} \bibinfo{year}{2019}\natexlab{}.
\newblock \showarticletitle{IoTGuard: Dynamic Enforcement of Security and
  Safety Policy in Commodity IoT.}. In \bibinfo{booktitle}{\emph{NDSS}}.
\newblock


\bibitem[Chen et~al\mbox{.}(2018)]%
        {iotfuzzer}
\bibfield{author}{\bibinfo{person}{Jiongyi Chen}, \bibinfo{person}{Wenrui
  Diao}, \bibinfo{person}{Qingchuan Zhao}, \bibinfo{person}{Chaoshun Zuo},
  \bibinfo{person}{Zhiqiang Lin}, \bibinfo{person}{XiaoFeng Wang},
  \bibinfo{person}{Wing~Cheong Lau}, \bibinfo{person}{Menghan Sun},
  \bibinfo{person}{Ronghai Yang}, {and} \bibinfo{person}{Kehuan Zhang}.}
  \bibinfo{year}{2018}\natexlab{}.
\newblock \showarticletitle{IoTFuzzer: Discovering Memory Corruptions in IoT
  Through App-based Fuzzing.}. In \bibinfo{booktitle}{\emph{NDSS}}.
\newblock


\bibitem[Chi et~al\mbox{.}(2020)]%
        {chi2020crossapp}
\bibfield{author}{\bibinfo{person}{Haotian Chi}, \bibinfo{person}{Qiang Zeng},
  \bibinfo{person}{Xiaojiang Du}, {and} \bibinfo{person}{Jiaping Yu}.}
  \bibinfo{year}{2020}\natexlab{}.
\newblock \showarticletitle{Cross-app interference threats in smart homes:
  Categorization, detection and handling}. In \bibinfo{booktitle}{\emph{2020
  50th Annual IEEE/IFIP International Conference on Dependable Systems and
  Networks (DSN)}}. IEEE, \bibinfo{pages}{411--423}.
\newblock


\bibitem[Ding et~al\mbox{.}(2021)]%
        {ding2021iotsafe}
\bibfield{author}{\bibinfo{person}{Wenbo Ding}, \bibinfo{person}{Hongxin Hu},
  {and} \bibinfo{person}{Long Cheng}.} \bibinfo{year}{2021}\natexlab{}.
\newblock \showarticletitle{IOTSAFE: Enforcing safety and security policy with
  real IoT physical interaction discovery}. In \bibinfo{booktitle}{\emph{the
  28th Network and Distributed System Security Symposium (NDSS 2021)}}.
\newblock


\bibitem[Dong et~al\mbox{.}(2023)]%
        {285062}
\bibfield{author}{\bibinfo{person}{Wei Dong}, \bibinfo{person}{Borui Li},
  \bibinfo{person}{Haoyu Li}, \bibinfo{person}{Hao Wu}, \bibinfo{person}{Kaijie
  Gong}, \bibinfo{person}{Wenzhao Zhang}, {and} \bibinfo{person}{Yi Gao}.}
  \bibinfo{year}{2023}\natexlab{}.
\newblock \showarticletitle{{LinkLab} 2.0: A Multi-tenant Programmable {IoT}
  Testbed for Experimentation with {Edge-Cloud} Integration}. In
  \bibinfo{booktitle}{\emph{20th USENIX Symposium on Networked Systems Design
  and Implementation (NSDI 23)}}. \bibinfo{publisher}{USENIX Association},
  \bibinfo{address}{Boston, MA}, \bibinfo{pages}{1683--1699}.
\newblock
\showISBNx{978-1-939133-33-5}
\urldef\tempurl%
\url{https://www.usenix.org/conference/nsdi23/presentation/dong}
\showURL{%
\tempurl}


\bibitem[Fernandes et~al\mbox{.}(2016)]%
        {fernandes2016smart}
\bibfield{author}{\bibinfo{person}{E. Fernandes}, \bibinfo{person}{J. Jung},
  {and} \bibinfo{person}{A. Prakash}.} \bibinfo{year}{2016}\natexlab{}.
\newblock \showarticletitle{Security Analysis of Emerging Smart Home
  Applications}. In \bibinfo{booktitle}{\emph{2016 IEEE Symposium on Security
  and Privacy (S\&P)}}. \bibinfo{publisher}{IEEE}.
\newblock
\showISSN{2375-1207}


\bibitem[Google(2018)]%
        {androidMonkey}
\bibfield{author}{\bibinfo{person}{Inc. Google}.}
  \bibinfo{year}{2018}\natexlab{}.
\newblock \bibinfo{booktitle}{}.
\newblock
\urldef\tempurl%
\url{https://developer.android.com/studio/test/other-testing-tools/monkey}
\showURL{%
\tempurl}


\bibitem[Goyal et~al\mbox{.}(2022)]%
        {goyal2022securing}
\bibfield{author}{\bibinfo{person}{Gaurav Goyal}, \bibinfo{person}{Peng Liu},
  {and} \bibinfo{person}{Shamik Sural}.} \bibinfo{year}{2022}\natexlab{}.
\newblock \showarticletitle{Securing Smart Home IoT Systems with
  Attribute-Based Access Control}. In \bibinfo{booktitle}{\emph{Proceedings of
  the 2022 ACM Workshop on Secure and Trustworthy Cyber-Physical Systems}}.
  \bibinfo{pages}{37--46}.
\newblock


\bibitem[Jia et~al\mbox{.}(2021)]%
        {10.1145/3460120.3484592}
\bibfield{author}{\bibinfo{person}{Yan Jia}, \bibinfo{person}{Bin Yuan},
  \bibinfo{person}{Luyi Xing}, \bibinfo{person}{Dongfang Zhao},
  \bibinfo{person}{Yifan Zhang}, \bibinfo{person}{XiaoFeng Wang},
  \bibinfo{person}{Yijing Liu}, \bibinfo{person}{Kaimin Zheng},
  \bibinfo{person}{Peyton Crnjak}, \bibinfo{person}{Yuqing Zhang},
  \bibinfo{person}{Deqing Zou}, {and} \bibinfo{person}{Hai Jin}.}
  \bibinfo{year}{2021}\natexlab{}.
\newblock \showarticletitle{Who's In Control? On Security Risks of Disjointed
  IoT Device Management Channels}. In \bibinfo{booktitle}{\emph{Proceedings of
  the 2021 ACM SIGSAC Conference on Computer and Communications Security}}
  (Virtual Event, Republic of Korea) \emph{(\bibinfo{series}{CCS '21})}.
  \bibinfo{publisher}{Association for Computing Machinery},
  \bibinfo{address}{New York, NY, USA}, \bibinfo{pages}{1289–1305}.
\newblock
\showISBNx{9781450384544}
\urldef\tempurl%
\url{https://doi.org/10.1145/3460120.3484592}
\showDOI{\tempurl}


\bibitem[Jia et~al\mbox{.}(2017)]%
        {contexiot17}
\bibfield{author}{\bibinfo{person}{Yunhan~Jack Jia}, \bibinfo{person}{Qi~Alfred
  Chen}, \bibinfo{person}{Shiqi Wang}, \bibinfo{person}{Amir Rahmati},
  \bibinfo{person}{Earlence Fernandes}, \bibinfo{person}{Z.~Morley Mao}, {and}
  \bibinfo{person}{Atul Prakash}.} \bibinfo{year}{2017}\natexlab{}.
\newblock \showarticletitle{{ContexIoT: Towards Providing Contextual Integrity
  to Appified IoT Platforms}}. In \bibinfo{booktitle}{\emph{21st Network and
  Distributed Security Symposium (NDSS)}}.
\newblock


\bibitem[Landi({[n.\,d.]})]%
        {iotsafety2019}
\bibfield{author}{\bibinfo{person}{Heather Landi}.}
  \bibinfo{year}{[n.\,d.]}\natexlab{}.
\newblock \bibinfo{title}{82\% of healthcare organizations have experienced an
  IoT-focused cyberattack, survey finds}.
\newblock
  \bibinfo{howpublished}{{https://www.fiercehealthcare.com/tech/82-healthcare-organizations-have-experienced-iot-focused-cyber-attack-survey-finds}}.
\newblock


\bibitem[Lee et~al\mbox{.}(2017)]%
        {lee2017fact}
\bibfield{author}{\bibinfo{person}{Sanghak Lee}, \bibinfo{person}{Jiwon Choi},
  \bibinfo{person}{Jihun Kim}, \bibinfo{person}{Beumjin Cho},
  \bibinfo{person}{Sangho Lee}, \bibinfo{person}{Hanjun Kim}, {and}
  \bibinfo{person}{Jong Kim}.} \bibinfo{year}{2017}\natexlab{}.
\newblock \showarticletitle{FACT: Functionality-centric access control system
  for IoT programming frameworks}. In \bibinfo{booktitle}{\emph{Proceedings of
  the 22nd ACM on Symposium on Access Control Models and Technologies}}. ACM.
\newblock


\bibitem[Manandhar et~al\mbox{.}(2020)]%
        {manandhar2020Helion}
\bibfield{author}{\bibinfo{person}{Sunil Manandhar}, \bibinfo{person}{Kevin
  Moran}, \bibinfo{person}{Kaushal Kafle}, \bibinfo{person}{Ruhao Tang},
  \bibinfo{person}{Denys Poshyvanyk}, {and} \bibinfo{person}{Adwait Nadkarni}.}
  \bibinfo{year}{2020}\natexlab{}.
\newblock \showarticletitle{Towards a natural perspective of smart homes for
  practical security and safety analyses}. In \bibinfo{booktitle}{\emph{2020
  IEEE Symposium on Security and Privacy (S\&P)}}. IEEE.
\newblock


\bibitem[Mazhar et~al\mbox{.}(2023)]%
        {mazhar2023maverick}
\bibfield{author}{\bibinfo{person}{M~Hammad Mazhar}, \bibinfo{person}{Li Li},
  \bibinfo{person}{Endadul Hoque}, {and} \bibinfo{person}{Omar Chowdhury}.}
  \bibinfo{year}{2023}\natexlab{}.
\newblock \showarticletitle{MAVERICK: An App-independent and Platform-agnostic
  Approach to Enforce Policies in IoT Systems at Runtime}. In
  \bibinfo{booktitle}{\emph{Proceedings of the 13th ACM Conference on Security
  and Privacy in Wireless and Mobile Networks 2020 (WiSec '20)}}.
\newblock


\bibitem[Munoz et~al\mbox{.}(2019)]%
        {munoz:hal-02266558}
\bibfield{author}{\bibinfo{person}{Jonathan Munoz}, \bibinfo{person}{Fabian
  Rincon}, \bibinfo{person}{Tengfei Chang}, \bibinfo{person}{Xavier
  Vilajosana}, \bibinfo{person}{Brecht Vermeulen}, \bibinfo{person}{Thijs
  Walcarius}, \bibinfo{person}{Wim van~de Meerssche}, {and}
  \bibinfo{person}{Thomas Watteyne}.} \bibinfo{year}{2019}\natexlab{}.
\newblock \showarticletitle{{OpenTestBed: Poor Man's IoT Testbed}}. In
  \bibinfo{booktitle}{\emph{{IEEE INFOCOM - CNERT : Workshop on Computer and
  Networking Experimental Research using Testbeds}}}. \bibinfo{address}{Paris,
  France}.
\newblock
\urldef\tempurl%
\url{https://inria.hal.science/hal-02266558}
\showURL{%
\tempurl}


\bibitem[Nafis et~al\mbox{.}(2023)]%
        {vetiotCNS2023}
\bibfield{author}{\bibinfo{person}{Akib~Jawad Nafis}, \bibinfo{person}{Omar
  Chowdhury}, {and} \bibinfo{person}{Endadul Hoque}.}
  \bibinfo{year}{2023}\natexlab{}.
\newblock \showarticletitle{VetIoT: On Vetting IoT Defenses Enforcing Policies
  at Runtime}. In \bibinfo{booktitle}{\emph{2023 IEEE Conference on
  Communications and Network Security (CNS)}}. \bibinfo{pages}{1--9}.
\newblock
\urldef\tempurl%
\url{https://doi.org/10.1109/CNS59707.2023.10288667}
\showDOI{\tempurl}


\bibitem[Nguyen et~al\mbox{.}(2018)]%
        {nguyen2018iotsan}
\bibfield{author}{\bibinfo{person}{Dang~Tu Nguyen}, \bibinfo{person}{Chengyu
  Song}, \bibinfo{person}{Zhiyun Qian}, \bibinfo{person}{Srikanth~V
  Krishnamurthy}, \bibinfo{person}{Edward~JM Colbert}, {and}
  \bibinfo{person}{Patrick McDaniel}.} \bibinfo{year}{2018}\natexlab{}.
\newblock \showarticletitle{IotSan: fortifying the safety of IoT systems}. In
  \bibinfo{booktitle}{\emph{Proceedings of the 14th International Conference on
  emerging Networking EXperiments and Technologies}}. ACM.
\newblock


\bibitem[openHAB(2019)]%
        {OpenHab}
\bibfield{author}{\bibinfo{person}{openHAB}.} \bibinfo{year}{2019}\natexlab{}.
\newblock \bibinfo{booktitle}{}.
\newblock
\urldef\tempurl%
\url{https://www.openhab.org}
\showURL{%
\tempurl}


\bibitem[Rahmati et~al\mbox{.}(2018)]%
        {rahmati2018tyche}
\bibfield{author}{\bibinfo{person}{Amir Rahmati}, \bibinfo{person}{Earlence
  Fernandes}, \bibinfo{person}{Kevin Eykholt}, {and} \bibinfo{person}{Atul
  Prakash}.} \bibinfo{year}{2018}\natexlab{}.
\newblock \showarticletitle{Tyche: A Risk-Based Permission Model for Smart
  Homes}. In \bibinfo{booktitle}{\emph{2018 IEEE Cybersecurity Development
  (SecDev)}}. IEEE.
\newblock


\bibitem[Redini et~al\mbox{.}(2021)]%
        {redini2021diane}
\bibfield{author}{\bibinfo{person}{Nilo Redini}, \bibinfo{person}{Andrea
  Continella}, \bibinfo{person}{Dipanjan Das}, \bibinfo{person}{Giulio
  De~Pasquale}, \bibinfo{person}{Noah Spahn}, \bibinfo{person}{Aravind
  Machiry}, \bibinfo{person}{Antonio Bianchi}, \bibinfo{person}{Christopher
  Kruegel}, {and} \bibinfo{person}{Giovanni Vigna}.}
  \bibinfo{year}{2021}\natexlab{}.
\newblock \showarticletitle{DIANE: Identifying Fuzzing Triggers in Apps to
  Generate Under-constrained Inputs for IoT Devices}. In
  \bibinfo{booktitle}{\emph{42nd IEEE Symposium on Security and Privacy 2021}}.
\newblock


\bibitem[Sikder et~al\mbox{.}(2020)]%
        {sikder2020kratos}
\bibfield{author}{\bibinfo{person}{Amit~Kumar Sikder},
  \bibinfo{person}{Leonardo Babun}, \bibinfo{person}{Z~Berkay Celik},
  \bibinfo{person}{Abbas Acar}, \bibinfo{person}{Hidayet Aksu},
  \bibinfo{person}{Patrick McDaniel}, \bibinfo{person}{Engin Kirda}, {and}
  \bibinfo{person}{A~Selcuk Uluagac}.} \bibinfo{year}{2020}\natexlab{}.
\newblock \showarticletitle{Kratos: Multi-user multi-device-aware access
  control system for the smart home}. In \bibinfo{booktitle}{\emph{Proceedings
  of the 13th ACM Conference on Security and Privacy in Wireless and Mobile
  Networks}}. \bibinfo{pages}{1--12}.
\newblock


\bibitem[SmartThings(2019)]%
        {smartThings}
\bibfield{author}{\bibinfo{person}{SmartThings}.}
  \bibinfo{year}{2019}\natexlab{}.
\newblock \bibinfo{booktitle}{}.
\newblock
\urldef\tempurl%
\url{https://www.smartthings.com/}
\showURL{%
\tempurl}


\bibitem[Sáez-de Cámara et~al\mbox{.}(2023)]%
        {10049670}
\bibfield{author}{\bibinfo{person}{Xabier Sáez-de Cámara},
  \bibinfo{person}{Jose~Luis Flores}, \bibinfo{person}{Cristóbal Arellano},
  \bibinfo{person}{Aitor Urbieta}, {and} \bibinfo{person}{Urko Zurutuza}.}
  \bibinfo{year}{2023}\natexlab{}.
\newblock \showarticletitle{Gotham Testbed: A Reproducible IoT Testbed for
  Security Experiments and Dataset Generation}.
\newblock \bibinfo{journal}{\emph{IEEE Transactions on Dependable and Secure
  Computing}} (\bibinfo{year}{2023}), \bibinfo{pages}{1--18}.
\newblock
\urldef\tempurl%
\url{https://doi.org/10.1109/TDSC.2023.3247166}
\showDOI{\tempurl}


\bibitem[Tian et~al\mbox{.}(2017)]%
        {tian2017smartauth}
\bibfield{author}{\bibinfo{person}{Yuan Tian}, \bibinfo{person}{Nan Zhang},
  \bibinfo{person}{Yueh-Hsun Lin}, \bibinfo{person}{XiaoFeng Wang},
  \bibinfo{person}{Blase Ur}, \bibinfo{person}{Xianzheng Guo}, {and}
  \bibinfo{person}{Patrick Tague}.} \bibinfo{year}{2017}\natexlab{}.
\newblock \showarticletitle{SmartAuth: User-Centered Authorization for the
  Internet of Things}. In \bibinfo{booktitle}{\emph{26th {USENIX} Security
  Symposium ({USENIX} Security 17)}}.
\newblock


\bibitem[Wang et~al\mbox{.}(2019)]%
        {wang2019charting}
\bibfield{author}{\bibinfo{person}{Qi Wang}, \bibinfo{person}{Pubali Datta},
  \bibinfo{person}{Wei Yang}, \bibinfo{person}{Si Liu}, \bibinfo{person}{Adam
  Bates}, {and} \bibinfo{person}{Carl~A Gunter}.}
  \bibinfo{year}{2019}\natexlab{}.
\newblock \showarticletitle{Charting the attack surface of trigger-action IoT
  platforms}. In \bibinfo{booktitle}{\emph{Proceedings of the 2019 ACM SIGSAC
  conference on computer and communications security}}.
  \bibinfo{pages}{1439--1453}.
\newblock


\bibitem[Wang et~al\mbox{.}(2018)]%
        {wang2018fear}
\bibfield{author}{\bibinfo{person}{Qi Wang}, \bibinfo{person}{Wajih~Ul Hassan},
  \bibinfo{person}{Adam Bates}, {and} \bibinfo{person}{Carl Gunter}.}
  \bibinfo{year}{2018}\natexlab{}.
\newblock \showarticletitle{Fear and Logging in the Internet of Things}. In
  \bibinfo{booktitle}{\emph{ISOC NDSS}}.
\newblock


\bibitem[Yahyazadeh et~al\mbox{.}(2020)]%
        {yahyazadeh2020patriot}
\bibfield{author}{\bibinfo{person}{Moosa Yahyazadeh},
  \bibinfo{person}{Syed~Rafiul Hussain}, \bibinfo{person}{Endadul Hoque}, {and}
  \bibinfo{person}{Omar Chowdhury}.} \bibinfo{year}{2020}\natexlab{}.
\newblock \showarticletitle{PatrIoT: Policy Assisted Resilient Programmable IoT
  System}. In \bibinfo{booktitle}{\emph{International Conference on Runtime
  Verification}}. Springer.
\newblock


\bibitem[Yahyazadeh et~al\mbox{.}(2019)]%
        {yahyazadeh2019expat}
\bibfield{author}{\bibinfo{person}{Moosa Yahyazadeh}, \bibinfo{person}{Proyash
  Podder}, \bibinfo{person}{Endadul Hoque}, {and} \bibinfo{person}{Omar
  Chowdhury}.} \bibinfo{year}{2019}\natexlab{}.
\newblock \showarticletitle{EXPAT: Expectation-based policy analysis and
  enforcement for appified smart-home platforms}. In
  \bibinfo{booktitle}{\emph{Proceedings of the 24th ACM Symposium on Access
  Control Models and Technologies}}.
\newblock


\bibitem[Yu and Chen(2020)]%
        {9290797}
\bibfield{author}{\bibinfo{person}{Keyang Yu} {and} \bibinfo{person}{Dong
  Chen}.} \bibinfo{year}{2020}\natexlab{}.
\newblock \showarticletitle{SmartAttack: Open-source Attack Models for Enabling
  Security Research in Smart Homes}. In \bibinfo{booktitle}{\emph{2020 11th
  International Green and Sustainable Computing Workshops (IGSC)}}.
  \bibinfo{pages}{1--8}.
\newblock
\urldef\tempurl%
\url{https://doi.org/10.1109/IGSC51522.2020.9290797}
\showDOI{\tempurl}


\end{thebibliography}

\appendix

\section{IoTGuard Re-Implementation}
\label{sec:appen:iotguard}
\iotguard \cite{iotguard2019ndss} is a defense mechanism built to protect IoT system from malicious IoT applications
(hereafter, apps) and unintended interactions among seemingly benign IoT applications. 
To ensure safety and security, 
\iotguard enforces policies on IoT applications at runtime. 

Policy enforcement mechanism of \iotguard is built using three modules: 
\textit{code instrumentor}, \textit{data collector}, and \textit{security service}.
To the best of our knowledge, implementations of data collector and security service are not publicly available.
\iotguard's code instrumentor was built for the \smt platform.
Since \sysname is developed for \oh, we could not reuse the publicly available code instrumentor of \iotguard.\footnote{\code{https://github.com/Beerkay/SmartAppAnalysis}}
Therefore, we had to re-implement all three modules of \iotguard for \oh. 

The purpose of the code instrumentor module of \iotguard is to instrument an IoT app
with necessary hooks, which guard each action to be taken by the app. 
At runtime, when an instrumented app is triggered and starts executing, these hooks
enable 
\iotguard's data collector to collect necessary data and 
invoke security service to verify safety and security of taking the contemplated action.
According to \cite{iotguard2019ndss}, 
when an app is about to take multiple actions in the same context, 
the code instrumentor must deploy one common hook for all such actions.
In our implementation of the code instrumentor, we first parsed the given app for \oh 
and inserted those policy enforcing hooks before the action statement of the app.
Our instrumentor follows the design mentioned in \cite{iotguard2019ndss}.

\iotguard's data collector collects necessary information (\ie, the occurred events and the contemplated actions) 
from the instrumented IoT apps at runtime and stores this information 
in a dynamic model consisting of states and transitions.
Each state represents an attribute of a device. 
Each transition from one state to another state represents the condition under which the state change has occurred. 
At a program level, this dynamic model is essentially a mutable directed graph where nodes represent states and 
edges denotes transitions.
For example, consider an IoT app named \code{light-control:} ``\code{when motion-active after sunset, turn on light.}''
Upon receiving data from the instrumented \code{light-control} app, 
the data collector will create a dynamic model which has an event node ``\code{motion=ACTIVE}'' 
and an action node ``\code{Light=ON}'' along with an edge between them with the condition ``\code{after sunset}''
(see \FIGREF~\ref{fig:light-app}).

\begin{figure}[!t]
	\centering
	\begin{subfigure}[c]{0.35\linewidth}
		\centering
		\includegraphics[width=0.8\textwidth]{\FIGLOC/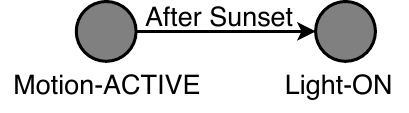}
		\caption{}
\label{fig:light-app}
	\end{subfigure}
	\begin{subfigure}[c]{0.6\linewidth}
		\centering
		\includegraphics[width=0.8\textwidth]{\FIGLOC/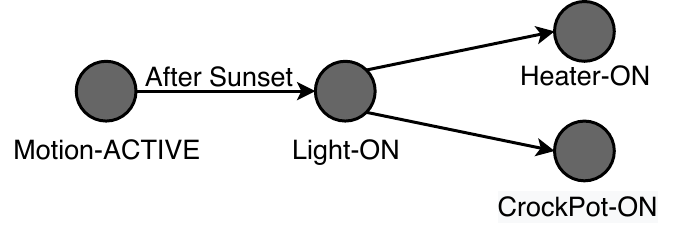}
\caption{}
		\label{fig:combined-model}
	\end{subfigure}
\caption{(a) A dynamic model of an IoT app and (b) the unified dynamic model of interacting IoT apps}
	\label{fig:dynamic-model}
	\vspace*{-1em}
\end{figure}

If multiple applications interact with each other, the data collector will create a unified dynamic model.
Consider this example. If the ``\code{Light=ON}'' event due to the action from the \code{light-control} app triggers 
another app, say, 
\code{heat-on} app: ``\code{Turn on the heater and crockpot when light is on.}'', 
the dynamic models of the \code{light-control} app and the \code{heat-on} app will be merged to generate a unified
dynamic model (see \FIGREF~\ref{fig:combined-model}).
In our implementation, we developed the data collector module using the Python's Networkx library and 
ensured this module has all the features of the original design \cite{iotguard2019ndss}.

\iotguard's security service module reads safety and security policies, enforces those policies on the dynamic models 
generated by the data collector, and conveys the result of policy enforcement to the instrumented IoT apps.
After receiving the data (\ie, event, action) from an instrumented IoT app, 
the data collector updates (or create) the dynamic model and invokes 
security service to enforce policies on the latest dynamic model.

\iotguard \cite{iotguard2019ndss} supports three types of policies: 
General Policies, Application Specific Policies, and Trigger-Action Specific Policies.
General policies are enforced directly on the dynamic model by applying graph-based algorithms 
(\eg, an cycle detection algorithm in the graph).
Application specific policies are written in a specific policy language described in \cite{iotguard2019ndss}.
We implemented a policy parser to read policies written in the \iotguard's policy language. 
To enforce application specific policy, our implementation of security service followed the description provided in \cite{iotguard2019ndss}. 

Each application specific policy is essentially a logical implication between a 
premise and a conclusion.
A premise/conclusion is a logical expression consisting of one or more atoms combined
using logical connectives (\eg, $\wedge$),  
where each atom denotes a proposition on a device attribute (\eg, comparing a device attribute
with a constant using a relational operator). Recall that device attributes (\eg, \code{Light=ON}) are represented as nodes in the dynamic model.
To enforce an application specific policy, the security service searches for a path in the dynamic model
-- from the node that satisfies the premise to a node that satisfies the conclusion.
If there exists such a path in the dynamic model and if it is a ``restrict'' policy, the security service labels it as a policy violation. 
Then, the security service removes the nodes that were recently added by the data collector 
prior to this round of policy enforcement
and returns a ``deny'' response to the instrumented IoT app.

To enforce a trigger-action specific policy, we tagged each physical device of the testbed as \code{trusted} and 
\code{secure} and all virtual triggers (\eg, {EmailSent}, {GoogleAssistantActivated}) as \code{untrusted} and 
\code{insecure}. When enforcing a trigger-action specific policy, the security service searches 
the dynamic model
for a path from a node with the \code{untrusted}/\code{insecure} tag to a node with 
the \code{trusted}/\code{secure} tag.
If there exists such a path, the security service denies the contemplated actions and 
removes the recently added nodes from the model.

\end{document}

 \typeout{get arXiv to do 4 passes: Label(s) may have changed. Rerun}